\documentclass{amsart}

\usepackage{latexsym}
\usepackage{amssymb}
\usepackage{mathrsfs}
\theoremstyle{plain}
\newtheorem{thm}{Theorem}
\newtheorem{prop}{Proposition}
\newtheorem{lemma}{Lemma}

\theoremstyle{definition}
\newtheorem{definition}{Definition}

\renewcommand{\Sp}{\ensuremath{\Sigma_{+}}}
\newcommand{\Sm}{\ensuremath{\Sigma_{-}}}
\newcommand{\No}{\ensuremath{N_{1}}}
\newcommand{\Nt}{\ensuremath{N_{2}}}
\newcommand{\Nth}{\ensuremath{N_{3}}}

\newcommand{\tx}{\ensuremath{\tilde{x}}}
\newcommand{\ty}{\ensuremath{\tilde{y}}}
\newcommand{\tbx}{\ensuremath{\tilde{\mathbf{x}}}}

\newcommand{\tg}{\ensuremath{\tilde{g}}}
\begin{document}
\title{The future asymptotics of Bianchi VIII vacuum solutions}
\author{Hans Ringstr\"{o}m}
\address{Max-Planck-Institut f\"{u}r Gravitationsphysik, Am M\"{u}hlenberg 1,
D-14476 Golm, Germany}

\begin{abstract}
Bianchi VIII vacuum solutions to Einstein's equations are causally
geodesically complete to the future, given an appropriate
time orientation, and the objective of this article is to analyze
the asymptotic behaviour of solutions in this time direction.
For the Bianchi class A spacetimes, there is a formulation
of the field equations that was presented in an article
by Wainwright and Hsu, and we will analyze
the asymptotic behaviour of solutions in these variables. 
We also try to give the analytic results a
geometric interpretation by analyzing how
a normalized version of the Riemannian metric on the spatial
hypersurfaces of homogeneity evolves.
\end{abstract}
\maketitle

\section{Introduction}
Let us begin by defining what we mean by Bianchi class A spacetimes.
We do this by defining the concept class A vacuum initial data.
Let us introduce some terminology. Let $G$ be a $3$-dimensional Lie
group, $e_{i}$, $i=1,2,3$ be a basis of the Lie algebra with structure
constants determined by  $[e_{i},e_{j}]=\gamma_{ij}^{k}e_{k}$. If
$\gamma_{ik}^{k}=0$, then the Lie algebra and Lie group are said to 
be of class A and  
\begin{equation}\label{eq:sconstants}
\gamma_{ij}^{k}=\epsilon_{ijm}n^{km}
\end{equation}
where the symmetric matrix $n^{ij}$ is given by
\begin{equation}\label{eq:ndef}
n^{ij}=\frac{1}{2}\gamma^{(i}_{kl}\epsilon_{}^{j)kl}.
\end{equation}
\begin{definition}\label{def:data}
\textit{Class A vacuum initial data} for Einstein's equations consist of
the following. A  Lie group $G$ of class A, a left invariant metric
$g$ on $G$ and a left invariant symmetric covariant two-tensor $k$ on
$G$ satisfying
\begin{equation}\label{eq:con1}
R_{g}-k_{ij}k^{ij}+(\mathrm{tr}_{g} k)^2=0
\end{equation}
and
\begin{equation}\label{eq:con2}
\nabla_{i}\mathrm{tr}_{g} k-\nabla^{j}k_{ij}=0
\end{equation}
where $\nabla$ is the Levi-Civita connection of $g$ and $R_{g}$ is
the corresponding scalar curvature, indices are raised and lowered
by $g$.
\end{definition}
Consider class A vacuum initial data.
We can choose a left invariant orthonormal basis $\{ e_{i}\}$ with 
respect to $g$ so that the corresponding matrix $n^{ij}$ defined in 
(\ref{eq:ndef}) is diagonal with diagonal elements $n_{1}$, $n_{2}$ 
and $n_{3}$. By an appropriate choice of orthonormal basis 
$n_{1}, n_{2}, n_{3}$ can be assumed to belong to one and only
one of the types given in table \ref{table:bianchiA}. We 
assign a Bianchi type to the initial data accordingly. 
This division constitutes a classification of the class A Lie
algebras, see Lemma \ref{lemma:liealg}.

We can divide the initial data into three classes,
\begin{enumerate}
\item Bianchi IX initial data. In this case the maximal globally hyperbolic 
  development of the initial data is future and past causally geodesically 
  incomplete. It expands for a certain period of time, reaches a 
  moment of maximal expansion and then recollapses. 
\item Bianchi I and VI$\mathrm{I}_{0}$ initial data
  with $\mathrm{tr}_{g}k=0$. The corresponding developments are 
  causally geodesically complete.
\item The remaining types of initial data. With a suitable time 
  orientation such data evolve into a  future
  causally geodesically complete spacetime  
  expanding forever into the future. The development is however
  past causally geodesically incomplete.
\end{enumerate}

In this paper we are interested in the third class of initial
data.

\begin{table}\label{table:bianchiA}
\caption{Bianchi class A.}
\begin{tabular}{@{}lccc}
Type & $n_{1}$ & $n_{2}$ & $n_{3}$ \\
I                   & 0 & 0 & 0 \\
II                  & + & 0 & 0 \\
V$\mathrm{I}_{0}$   & 0 & + & $-$ \\
VI$\mathrm{I}_{0}$  & 0 & + & + \\
VIII                & $-$ & + & + \\
IX                  & + & + & + \\
\end{tabular}
\end{table}

There is a formulation of Einstein's equations due to Ellis and
MacCallum \cite{emac} covering, among other things, the Bianchi
class A spacetimes. In an article by Wainwright and Hsu \cite{whsu}, a
normalized version of the corresponding variables together with a
different time coordinate were introduced. The main part of this
paper consists of an analysis of the asymptotic behaviour of these
variables. Our main interest is in Bianchi VIII, but we also consider
the other Bianchi class A spacetimes in the third class as above.
The Bianchi I, II and VI${}_{0}$ cases are quickly handled, and are 
included for completeness, 
but it takes some time to analyze the asymptotics of
Bianchi VII${}_{0}$. Observe that Bianchi VII${}_{0}$ perfect fluids
have been considered in \cite{uggla}.
Bianchi VIII is dealt with definitively, in the sense that it is
proven to what the variables of Wainwright and Hsu converge. 
However, in certain situations this may not necessarily be enough.
For instance, it is proven that $N_{2}+N_{3}$ converge to infinity,
if $N_{2},\ N_{3}>0$ and $N_{1}<0$, and that $\Nt-\Nth$ converge
to zero. It is conceivable that if one is interested in computing
curvature invariants for instance, one has to consider expressions
of the form $\Nt^2-\Nth^2$, concerning which the results contained
in this paper say nothing.

Let us also mention that even though it does not require much
time to sort out the asymptotics for Bianchi VI${}_{0}$, the results 
yield an interesting consequence. In fact, one gets the conclusion that
Bianchi VIII is very different from Bianchi IX as far as the behaviour
towards the singularity is concerned. In \cite{jag2} it was proven that
generic Bianchi IX solutions converge to an attractor, where by the 
attractor, we mean the set of vacuum type I and II points, and the
generic points are obtained by subtracting a finite number of positive
codimension submanifolds from the manifold of Bianchi type IX points.
It was also proven that for generic Bianchi IX solutions, the
convergence to the attractor is almost monotone in the following
sense. Given an $\epsilon>0$, there is a $\delta>0$ such that if $x$ 
constitutes Bianchi IX initial data closer to the attractor than
$\delta$, then applying the flow to $x$ in the past time direction
will not result in points further away from the attractor than 
$\epsilon$. By distance, we here mean the ordinary Euclidean distance
in the Wainwright Hsu variables in $\mathbb{R}^{6}$, the relevant space 
in the Bianchi IX matter case. This result, 
together with the result that generic Bianchi IX solutions have an
$\alpha$-limit point on the Kasner circle, yield the conclusion
that generic solutions converge to the attractor. The almost monotone
convergence constitutes the most difficult step. In Proposition
\ref{prop:wierd}, we prove that the corresponding statement for 
Bianchi VIII is not true. In fact, we prove that there is a fixed number 
$\eta>0$, such that for any $\epsilon>0$ we can find Bianchi VIII 
initial data closer
to the Kasner circle than $\epsilon$ which will at one point to the past 
be at a distance further than $\eta$ away from the attractor. This
does of course not prove that Bianchi VIII solutions may not converge to 
the attractor, but it does imply that the main part of the argument
concerning Bianchi IX presented in \cite{jag2} is of no help in proving
convergence to the attractor, and that a
completely different method is required, assuming one still
believes that generic Bianchi VIII solutions converge to the attractor
towards the singularity.

Let us try to explain the argument concerning the future asymptotics of
a Bianchi VIII solution with $\No<0$ and $\Nt,\ \Nth>0$.  By a
monotonicity argument, one can prove that $\No\rightarrow 0$
and $\Nt,\ \Nth\rightarrow \infty$ as $\tau\rightarrow \infty$.
This is a very important observation in that it introduces a natural
concept of order of magnitude; powers of $\Nt+\Nth$.
Writing the constraint as
\[
\Sp^2+\Sm^2+\frac{3}{4}[\No^2+(\Nt-\Nth)^2-2\No(\Nt+\Nth)]=1,
\]
we conclude that all the terms involved are bounded since they
are all non-negative. The derivatives of
$\Sp$ and $\No(\Nt+\Nth)$ are bounded by constants independent of
the initial data, but we have
\[
\tx'=-3(\Nt+\Nth)\ty+...,\
\ty'=3(\Nt+\Nth)\tx+...
\]
where $\tx=\Sm$ and $\ty=\sqrt{3}(\Nt-\Nth)/2$, and the dots represent
expressions that can be bounded by constants independent of 
which Bianchi VIII solution we are considering. Thus, if
$(\tx^2+\ty^2)^{1/2}$ is big in comparison with $(\Nt+\Nth)^{-1}$,
$\ty$ and $\tx$ will essentially behave as sine and cosine, and the
frequency of the oscillation will go to infinity as $\tau\rightarrow
\infty$. The oscillations in $\tx$ and $\ty$ are not so interesting,
but the evolution in $\Sp$ and $\No(\Nt+\Nth)$ is. For that reason,
it is natural to consider what happens from, say, one time at which
$\Sm=0$ to the next. Before doing so, some effort has to be put 
into estimating the error committed in approximating $\ty$ and 
$\tx$ with sine and cosine, into proving that there actually
are zeros of $\Sm$ and estimating the variation of different objects
during the time that elapses between the zeros. After having done that, 
it will be possible to express
$\Sp$ and $\No(\Nt+\Nth)$ at a zero of $\Sm$ in terms of the values
of the same expressions in the previous zero. In other words, it will
be possible to replace the continuous flow with a discrete map depending
on only two variables. This approximation will not always be valid, 
but when it is not, it can be replaced by less sophisticated 
arguments. The discrete map is then the main tool in a sequence
of technical arguments proving the desired result.

When we evolve initial data we get a Lorentz metric 
$\bar{g}=- d t^{2}+g(t)$ on $I\times G$ where $I$ is an open 
interval and $g(t)$ is a Riemannian left invariant metric on $G$. 
Let $\Gamma$ be a subgroup of the group of diffeomorphisms of $G$, 
acting properly discontinuously on $G$. We can consider $\Gamma$ 
as acting on $I\times G$ by letting $\Gamma$ act only on $G$. In 
situations where $g(t)$ is invariant under $\Gamma$, we can take the
quotient to get a solution to Einstein's equations on 
$I\times M$, where $M=G/\Gamma$. Assuming $M$ to be compact
we can define the reduced Hamiltonian, c.f. \cite{fimon},
\[
H_{\mathrm{reduced}}=-(\mathrm{tr}_{g}k)^{3}\mathrm{vol}(M,g),
\]
where $k(t)$ is the second fundamental form of $\{ t\}\times M$.
We can consider it to be a function of $t$.
This object is of interest in the context of the work by Fischer
and Moncrief \cite{fimon}. We prove that the reduced Hamiltonian
converges to zero for all the initial data in class three, though
it should be remarked that the spatial hypersurfaces
should be of Yamabe type -1, see \cite{fimon}, for
this result to be of interest. Finally, let us refer the
reader to \cite{bar} and references therein for further
discussion of Bianchi VIII vacuum solutions.

The equations of Wainwright and Hsu are to be found in section 
\ref{section:whsu}. We mention some properties and describe 
an important tool in the analysis of the asymptotics of solutions 
to these equations; the monotonicity principle. In section
\ref{section:metric} we relate the Lorentz metric on the development
with the Wainwright Hsu variables. A formula expressing the
reduced Hamiltonian in terms of these variables is also given.
The main part of this paper consists of an analysis of the asymptotic
behaviour of solutions to the equations of Wainwright and Hsu.
This analysis occupies the remaining sections except for the last.
The final section contains an analysis of the
asymptotic behaviour of the Riemannian metric on the spatial 
hypersurfaces of homogeneity.

\section{The equations of Wainwright and Hsu}\label{section:whsu}
We formulate the equations we will use here and state some 
properties. The following section contains a derivation. 
Einstein's vacuum equations take the 
following form in the formulation due to Wainwright and 
Hsu
\begin{eqnarray}
\No'&=&(q-4\Sp)\No  \nonumber \\
\Nt'&= &(q+2\Sp +2\sqrt{3}\Sm)\Nt  \nonumber \\
\Nth'&=& (q+2\Sp -2\sqrt{3}\Sm)\Nth  \label{eq:whsu}\\
\Sp'&=& -(2-q)\Sp-3S_{+}  \nonumber \\
\Sm'&=& -(2-q)\Sm-3S_{-} \nonumber 
\end{eqnarray}
where the prime denotes derivative with respect to $\tau$ and
\begin{eqnarray}
q & = & 2(\Sp^2+\Sm^2) \nonumber \\
S_{+} & = & \frac{1}{2}[(\Nt-\Nth)^2-\No(2\No-\Nt-\Nth)]
\label{eq:whsudef}\\
S_{-} & = & \frac{\sqrt{3}}{2}(\Nth-\Nt)(\No-\Nt-\Nth). \nonumber
\end{eqnarray}
The vacuum Hamiltonian constraint is
\begin{equation}
\Sp^2+\Sm^2+\frac{3}{4}[\No^2+\Nt^2+\Nth^2-2(\No\Nt+\Nt\Nth+\No\Nth)]=1.
\label{eq:constraint}
\end{equation}
The above equations have certain
symmetries described in Wainwright and Hsu \cite{whsu}. By permuting
$\No,\Nt,\Nth$ arbitrarily  we get new solutions if we at the same 
time carry out appropriate
combinations of rotations by integer multiples of $2\pi/3$
and reflections in the $(\Sp,\Sm)$-plane.
Changing the sign of all the $N_{i}$ at the same time does not change
the equations. Classify points $(\No,\Nt,\Nth,\Sp,\Sm)$ according to
the values of $\No,\Nt,\Nth$ in the same way as in table
\ref{table:bianchiA}. Since the sets 
$N_{i}>0$, $N_{i}<0$ and $N_{i}=0$ are invariant under the flow of 
the equation we may classify solutions to 
(\ref{eq:whsu})-(\ref{eq:constraint}) accordingly. When we speak of 
Bianchi VIII solutions we will assume two $N_{i}>0$ and one $<0$. 
The \textit{Kasner circle} is the subset of $\Sp\Sm \No\Nt\Nth$-space
where $N_{i}=0$ and $\Sp^2+\Sm^2=1$.

We only consider solutions to (\ref{eq:whsu})-(\ref{eq:constraint})
which are not of Bianchi IX type. 
By the constraint (\ref{eq:constraint}) we conclude that $q\leq
2$ for the entire solution. As a consequence the $N_{i}$ cannot 
grow faster than exponentially due to (\ref{eq:whsu}). The 
solutions we consider can thus not blow up in a finite time
so that we have existence intervals of the form $(-\infty,\infty)$.
\begin{definition}
Let $\mathcal{K}_{1},\ \mathcal{K}_{2}$ and $\mathcal{K}_{3}$ 
be defined as the subset of the Kasner circle
on which $q-4\Sp<0,\ q+2\Sp+2\sqrt{3}\Sm<0$ and
$q+2\Sp-2\sqrt{3}\Sm<0$ respectively, c.f. (\ref{eq:whsu}).
\end{definition}

The set $\Sm=0$, $\Nt=\Nth$ is invariant under the flow of
(\ref{eq:whsu})-(\ref{eq:constraint}). Applying the symmetries to 
this set we get new invariant sets.

The concepts $\alpha$- and $\omega$-limit set will be useful.

\begin{definition}\label{def:als}
Let $f\in C^{\infty}(\mathbb{R}^{n},\mathbb{R}^{n})$, and consider 
a solution $x$ to the equation
\begin{equation}\label{eq:dxdteqf}
\frac{ d x}{ d t}=f\circ x,\ x(0)=x_{0},
\end{equation}
with maximal existence interval $(t_{-},t_{+})$. 
We call a point $x_{*}$ an $\omega$-\textit{limit point} of the 
solution $x$, if there is a sequence $t_{k}\rightarrow
t_{+}$ with $x(t_{k})\rightarrow x_{*}$. The $\omega$-\textit{limit
set} of $x$ is the set of its $\omega$-limit points. The
$\alpha$-limit set is defined similarly by replacing $t_{+}$ with
$t_{-}$.
\end{definition}

\begin{lemma}\label{lemma:als}
Let $f$ and $x$ be as in Definition \ref{def:als}.
The $\omega$-limit set of $x$ is closed and invariant under the flow of
$f$. If there is a $T$ such that $x(t)$ is contained in a compact 
set for $t\geq T$, then the $\omega$-limit set of $x$ is connected.
\end{lemma}
\textit{Proof}. See e.g. \cite{irwin}. $\Box$

The following lemma will be a basic tool in the analysis of the
asymptotics, we will refer to it as \textit{the monotonicity
principle}.
\begin{lemma}\label{lemma:monotone}
Consider (\ref{eq:dxdteqf}).
Let $U$ be an open subset of $\mathbb{R}^{n}$ and $M$ a
closed subset which is invariant
under the flow of the vectorfield $f$. Assume there is a 
function $F\in C(U,\mathbb{R})$ such that $F(x(t))$ is strictly monotone
for any solution $x(t)$ of (\ref{eq:dxdteqf}), as long as 
$x(t)\in U\cap M$. Then no solution of (\ref{eq:dxdteqf}) whose
image is contained in $U\cap M$ has an $\alpha$- or $\omega$-limit point 
in $U$.
\end{lemma}
\textit{Remark}. Observe that one can use $M=\mathbb{R}^{n}$. We will
mainly choose $M$ to be the closed invariant submanifold of 
$\mathbb{R}^{5}$ defined by (\ref{eq:constraint}). If one $N_{i}$ is
zero and two are non-zero we consider the number of variables to
be four etc.

\textit{Proof}. Let $x$ be a solution to (\ref{eq:dxdteqf}) which is
contained in $U\cap M$, and which has maximal existence interval 
$(t_{-},t_{+})$. Then $F\circ x$ is 
strictly monotone. Suppose $p\in U$ is an $\omega$-limit point
of $x$, so that there is a sequence $t_{n}\rightarrow t_{+}$
such that $x(t_{n})\rightarrow p$. Thus 
$F(x(t_{n}))\rightarrow F(p)$, but $F\circ x$ is monotone
so that $F(x(t))\rightarrow F(p)$ as $t\rightarrow t_{+}-$. 
Thus $F(q)=F(p)$ for all
$\omega$-limit points $q$ of $x$. Since $M$ is closed $p\in M$.
The solution $\bar{x}$ of (\ref{eq:dxdteqf}) with initial value $p$
is contained in $M$ by the invariance property of $M$, it
consists of $\omega$-limit points of $x$ so that $F(\bar{x}(t))=F(p)$
which is constant. Furthermore, on an open set containing zero it 
takes values in $U$ contradicting the assumptions of the lemma.
The argument for the $\alpha$-limit set is similar.
$\Box$

\section{Relation between the variables of Wainwright and Hsu
and the space time metric}\label{section:metric}
We need to relate the evolution of the variables in
the formulation due to Wainwright and Hsu to the evolution
of the corresponding spacetime. In order to do this we
first derive the formulation first presented by Ellis and
MacCallum \cite{emac}. Then we derive equations 
(\ref{eq:whsu})-(\ref{eq:constraint}) as in \cite{whsu}.
Secondly, we turn the argument around and start with
solutions to the equations of Ellis and MacCallum and then
construct the spacetime metric. The second part uses the
first part. We can then express the spacetime metric in
terms of the variables of Wainwright and Hsu. Finally,
we find expressions for the reduced Hamiltonian in terms
of the Wainwright Hsu variables.

\subsection{Derivation of the equations of Wainwright and Hsu}
Consider first a special class of spatially homogeneous
four dimensional Lorentz manifolds of the form
\begin{equation}\label{eq:structure}
(I\times G,- d t^2+\chi_{ij}(t)\xi^{i}\otimes \xi^{j})
\end{equation}
where $I$ is an open interval, $G$ is a Lie group of class A, 
$\chi_{ij}$ is a smooth positive definite matrix and the $\xi^{i}$ are 
the duals of a left invariant basis on $G$. 

We express the condition that a manifold of the form
(\ref{eq:structure}) satisfies Einstein's vacuum equations 
in terms of the variables of Ellis and MacCallum. 
As these variables are defined in terms of
a suitable orthonormal basis, we begin by constructing it. 
Let  $e_{0}=\partial_{t}$ and $e_{i}=a_{i}^{\ j}Z_{j}$, 
i=1,2,3, be an orthonormal basis, where  $a$ is a $C^{\infty}$ 
matrix valued function of $t$  and the $Z_{i}$ are the duals of
$\xi^{i}$. Below, Latin indices will be raised and lowered by 
$\delta_{ij}$.

By the following argument we can assume
$<\nabla_{e_{0}}e_{i},e_{j}>=0$. Let the matrix $A$ satisfy 
$e_{0}(A)+AB=0$, $A(0)=\mathrm{Id}$ 
where $B_{ij}=<\nabla_{e_{0}}e_{i},e_{j}>$ and $\mathrm{Id}$ is the 
$3\times 3$ identity matrix. Then $A$ is smooth and $SO(3)$ valued 
and if $e_{i}'=A_{i}^{\ j}e_{j}$ then  
$<\nabla_{e_{0}}e_{i}',e_{j}'>=0$.

Let
\begin{equation}
\theta(X,Y)=<\nabla_{X}e_{0},Y>,
\end{equation}
$\theta_{\alpha \beta}=\theta(e_{\alpha},e_{\beta})$ and
$[e_{\beta},e_{\delta}]=\gamma_{\beta \delta}^{\alpha}e_{\alpha}$  
where Greek indices run from $0$ to $3$. The objects $\theta_{\alpha 
\beta}$ and $\gamma_{\beta \delta}^{\alpha}$ will be viewed as
smooth functions from $I$ to some suitable $\mathbb{R}^{k}$ and
the variables will be defined in terms of them.

Observe that $[Z_{i},e_{0}]=0$. The $e_{i}$ span the tangent space of
$G$ and $<[e_{0},e_{i}],e_{0}>=0$. 
We get $\theta_{00}=\theta_{0i}=0$ and $\theta_{\alpha\beta}$
symmetric. We also have $\gamma_{ij}^{0}=\gamma_{0i}^{0}=0$ and
$\gamma_{0j}^{i}=-\theta_{ij}$. We let $n$ be defined as in 
(\ref{eq:ndef}) and
\[
\sigma_{ij}=\theta_{ij}-\frac{1}{3}\theta \delta_{ij}
\]
where we by abuse of notation have written $\mathrm{tr}(\theta)$ as 
$\theta$. 

We compute the Einstein tensor in terms of $n$, $\sigma$ 
and $\theta$. The Jacobi identities for $e_{\alpha}$ yield
\begin{equation}\label{eq:dndt}
e_{0}(n_{ij})-2n_{k(i}^{}\sigma_{j)}^{\ k}+\frac{1}{3}\theta n_{ij}=0.
\end{equation}
The $0i$-components of the Einstein equations are equivalent to
\begin{equation}\label{eq:commute2}
\sigma_{i}^{\ k}n_{kj}-n_{i}^{\ k}\sigma_{kj}=0.
\end{equation}
Letting  
$b_{ij}=2n_{i}^{\ k}n_{kj}-\mathrm{tr}(n) n_{ij}$ and
$s_{ij}=b_{ij}-\frac{1}{3}\mathrm{tr}(b)\delta_{ij}$
the trace free part of the $ij$ equations are
\begin{equation}\label{eq:dsdt}
e_{0}(\sigma_{ij})+\theta\sigma_{ij}+s_{ij}=0.
\end{equation}
The fact that $R_{00}=0$ yields the Raychaudhuri equation
\begin{equation}\label{eq:raychaudhuri}
e_{0}(\theta)+\theta_{ij}\theta^{ij}=0
\end{equation}
and using this together with the trace of the $ij$-equations yields a
constraint
\begin{equation}\label{eq:constraint1}
\sigma_{ij}\sigma^{ij}+(n_{ij}n^{ij}-\frac{1}{2}\mathrm{tr}(n)^2)=
\frac{2}{3}\theta^2.
\end{equation}
Equations (\ref{eq:dndt})-(\ref{eq:constraint1}) are special cases
of equations given in Ellis and MacCallum \cite{emac}. 
At  $t=0$ we may diagonalize $n$ and $\sigma$ simultaneously
since they commute (\ref{eq:commute2}). Rotating $e_{\alpha}$ by the 
corresponding
element of $SO(3)$ yields upon going through the definitions that the
new $n$ and $\sigma$ are diagonal at $t=0$. Collect the 
off-diagonal terms of $n$ and $\sigma$ in one vector $v$. By 
(\ref{eq:dndt}) and (\ref{eq:dsdt}) there is a time dependent matrix
$C$ such that $\dot{v}=Cv$ so that $v(t)=0\ \forall t$ since 
$v(0)=0$. Since the rotation was time independent 
$<\nabla_{e_{0}}e_{i},e_{j}>=0$ holds in the new basis. 

Introduce, as in Wainwright and Hsu \cite{whsu},
$\Sigma_{ij}=\sigma_{ij}/\theta,\
N_{ij}=n_{ij}/\theta,\
B_{ij}=2N_{i}^{\ k}N_{kj}-N^{k}_{\ k}N_{ij}$ and 
$S_{ij}=B_{ij}-\frac{1}{3}B^{k}_{\ k}\delta_{ij}$
and define a new time coordinate $\tau$, independent of time
orientation, by
\begin{equation}\label{eq:dtdtau}
\frac{ d t}{ d\tau}=\frac{3}{\theta}.
\end{equation}
Let
$\Sp=\frac{3}{2}(\Sigma_{22}+\Sigma_{33})$ and 
$\Sm=\sqrt{3}(\Sigma_{22}-\Sigma_{33})/2$. If we let $N_{i}$ be the 
diagonal elements of $N_{ij}$, equations (\ref{eq:dndt}) and 
(\ref{eq:dsdt}) turn into (\ref{eq:whsu})-(\ref{eq:constraint})
The Raychaudhuri equation (\ref{eq:raychaudhuri}) takes the form
\begin{equation}\label{eq:raychaudhuri2}
\theta'=-(1+q)\theta.
\end{equation}

\subsection{Construction of the spacetime}
Above we derived Einstein's equations assuming the spacetime to be
of a special form. Next we construct a spacetime of the form 
(\ref{eq:structure}) beginning with initial data as in Definition
\ref{def:data}. The construction allows us to express the evolution
of the metric in terms of the variables of Wainwright and Hsu.
One step is the following lemma, which also classifies the class A
Lie algebras.

\begin{lemma}\label{lemma:liealg}
Table \ref{table:bianchiA} constitutes a classification of the class
A Lie algebras. Consider an arbitrary basis $\{e_{i}\}$ of the Lie 
algebra. Then by applying an orthogonal matrix to it, we can construct 
a basis $\{e_{i}'\}$ such that the corresponding $n'$ defined by
(\ref{eq:ndef}) has diagonal elements of one of the types given in
table \ref{table:bianchiA}.
\end{lemma}
\textit{Proof}. Let $e_{i}$ be a 
basis for the Lie algebra and $n$ be defined as in (\ref{eq:ndef}).
If we change the basis according to 
$e_{i}'=(A^{-1})_{i}^{\ j}e_{j}$ then $n$ transforms to
\begin{equation}\label{eq:transform}
n'=(\det A)^{-1}A^{t}nA
\end{equation}
Since $n$ is symmetric we assume from here on that the basis is such 
that it is diagonal. The matrix $A=\mathrm{diag}(1\ 1\ -1)$ changes 
the sign of $n$. A suitable orthogonal matrix performs even 
permutations of the diagonal. The number of non-zero elements on the 
diagonal is invariant under transformations (\ref{eq:transform})
taking one diagonal matrix to another. If $A=(a_{ij})$ and the 
diagonal matrix $n'$ is constructed as in (\ref{eq:transform}) we have
$n'_{kk}=(\det A)^{-1}\sum_{i=1}^{3}a_{ik}^{2}n_{ii}$ so that if all 
the diagonal elements of $n$ have the same sign the same is true
for $n'$. The statements of the lemma follow. $\Box$

Consider class A vacuum initial data $(G,
g,k)$ with notation as in Definition \ref{def:data}. 
We construct a spacetime of the form (\ref{eq:structure}) whose 
induced metric and second fundamental form on $\{ 0\}\times G$
is $g$ and $k$. Let $e_{i}'$, $i=1,2,3$ be a left invariant
orthonormal basis. We can assume the corresponding $n'$ to be of one
of the forms given in table \ref{table:bianchiA} by Lemma
\ref{lemma:liealg}. The content of (\ref{eq:con2}) is that 
$k_{ij}=k(e_{i}',e_{j}')$ and  $n'$ are to commute. We may thus also 
assume $k_{ij}$ to be diagonal without changing the earlier conditions
of the construction. If we let $n(0)=n'$, $\theta(0)=
-\mathrm{tr}_{g}k$ and $\sigma_{ij}(0)=-k_{ij}+\theta\delta_{ij}/3$ 
then (\ref{eq:con1}) is the same as (\ref{eq:constraint1}). Let $n$,
$\sigma$ and $\theta$ satisfy (\ref{eq:dndt}), (\ref{eq:dsdt}) and
(\ref{eq:raychaudhuri}) with initial values as specified above. 
Since (\ref{eq:constraint1}) is satisfied at $t=0$ it is satisfied 
for all times. For reasons given in connection with
(\ref{eq:constraint1}) $n$ and $\sigma$ will 
remain diagonal so that (\ref{eq:commute2}) will always hold. 

Let $M=I\times G$ where $I$ is the maximal existence interval for
solutions to (\ref{eq:dndt})-(\ref{eq:constraint1}) with initial data 
as above. We construct a basis $e_{\alpha}$ with the same properties 
as the basis used in deriving
(\ref{eq:dndt})-(\ref{eq:constraint1}). Then we define a metric by 
demanding that the basis be orthonormal and show that the
corresponding $\tilde{n}$, 
$\tilde{\sigma}$ and $\tilde{\theta}$ coincide with $n$, $\sigma$ 
and $\theta$. We will thereby have constructed a Lorentz manifold 
satisfying Einstein's vacuum equations with the correct initial 
conditions.

Let $n_{i}$ and $\sigma_{i}$ denote the diagonal elements of $n$ and
$\sigma$ respectively. Let $f_{i}(0)=1$ and
$\dot{f_{i}}/f_{i}=2\sigma_{i}-\theta/3$.
Let $a_{i}=(\Pi_{j\neq i}f_{j})^{1/2}$ and
define $e_{i}=a_{i}e_{i}'$. Then $\tilde{n}$ associated to $e_{i}$ 
equals $n$. We complete the basis by letting $e_{0}=\partial_{t}$. 
Define a metric $<\cdot,\cdot>$ on $M$ by demanding $e_{\alpha}$ to be 
orthonormal with $<e_{0},e_{0}>=-1$ and $<e_{i},e_{i}>=1$, $i=1,2,3$
and let $\nabla$ be the associated Levi-Civita
connection. 

Compute $<\nabla_{e_{0}}e_{i},e_{j}>=0$.
If $\tilde{\theta}(X,Y)=<\nabla_{X}e_{0},Y>$ and
$\tilde{\theta}_{\mu\nu}=\tilde{\theta}(e_{\mu},e_{\nu})$, then 
$\tilde{\theta}_{00}=\tilde{\theta}_{i0}=
\tilde{\theta}_{0i}=0$. Furthermore,
\[
\frac{1}{a_{j}}e_{0}(a_{j})\delta_{ij}=-\tilde{\theta}_{ij}
\]
(no summation over $j$) so that $\tilde{\theta}_{ij}$ is diagonal and
$\mathrm{tr}\tilde{\theta}=\theta$. Finally,
\[
-\tilde{\sigma}_{ii}=-\tilde{\theta}_{ii}+\frac{1}{3}\theta=-\sigma_{i}.
\]
The constructed Lorentz manifold thus satisfies Einstein's 
vacuum equations. By the above we have.

\begin{lemma}\label{lemma:development}
Given initial data as in Definition \ref{def:data}, we get a solution
to Einstein's equations
\begin{equation}\label{eq:metric}
\bar{g}=- d t^{2}+\sum_{i=1}^{3}a_{i}^{-2}(t)\xi^{i}\otimes \xi^{i}
\end{equation}
where $\xi^{i}$ are the duals of a basis of the Lie algebra of $G$ 
and
\begin{equation}\label{eq:aidef}
a_{i}(t)=\exp (-\int_{0}^{t}(\sigma_{i}+\frac{1}{3}\theta) d s)
\end{equation}
where $\theta,\ \sigma_{i}$ (and $n_{i}$) constitute solutions to 
(\ref{eq:dndt})-(\ref{eq:constraint1}) with suitable initial
conditions. If $I$ is the corresponding existence interval, the
metric (\ref{eq:metric}) is defined on the manifold $I\times G$.
\end{lemma}
By \cite{jag2} this metric is globally hyperbolic, future 
causally geodesically complete and $\tau\rightarrow \infty$ corresponds
to the geodesically complete direction, assuming the initial data are
not of type IX.

\subsection{The rescaled metric}
The equations (\ref{eq:metric}) and (\ref{eq:aidef}) express
the spacetime metric in terms of the Ellis-MacCallum variables.
Let $g(t)$ denote the 
Riemannian metric induced on $\{ t\}\times G$, which we will also
consider to be a metric on $G$. Define a rescaled version
of this metric by
\begin{equation}\label{eq:rescaled}
\tg(t)=(a_{1}a_{2}a_{3})^{2/3}(t)\sum_{i=1}^{3}a_{i}^{-2}(t)\xi^{i}\otimes
\xi^{i}.
\end{equation}
We can write it as
\begin{equation}\label{eq:rescaled2}
\tg(t)=\sum_{i=1}^{3}\lambda_{i}(t)\xi^{i}\otimes \xi^{i}
\end{equation}
where 
\[
\lambda_{i}(t)=\exp (\int_{0}^{t}2\sigma_{i} d s).
\]
Defining a new time coordinate by
\begin{equation}\label{eq:taudef}
\frac{ d t}{ d\tau}=\frac{3}{\theta},\ \tau(0)=0
\end{equation}
we get
\[
\int_{0}^{t}2\sigma_{i} d s=
\int_{0}^{\tau}2\sigma_{i}\frac{3}{\theta} d\tau'=
\int_{0}^{\tau}6\Sigma_{i} d\tau'.
\]
Thus if we consider the $\lambda_{i}$ to be functions of $\tau$
we have
\begin{equation}\label{eq:lidef}
\lambda_{i}(\tau)=\exp( \int_{0}^{\tau}6\Sigma_{i} d\tau').
\end{equation}
Similarly,
\begin{equation}\label{eq:aitau}
a_{i}(\tau)=\exp(-\int_{0}^{\tau}(3\Sigma_{i}+1) d s).
\end{equation}
However, the $\Sigma_{i}$ can be expressed in terms of $\Sp$ and
$\Sm$. We have
\begin{equation}\label{eq:sigi}
\Sigma_{1}=-\frac{2}{3}\Sp,\ 
\Sigma_{2}=\frac{1}{3}\Sp+\frac{1}{\sqrt{3}}\Sm,\
\Sigma_{3}=\frac{1}{3}\Sp-\frac{1}{\sqrt{3}}\Sm.
\end{equation}
Thus, if we know what $\Sp$ and $\Sm$ converge to as $\tau\rightarrow
\infty$, we roughly know how $\tg$ behaves asymptotically.

\subsection{The reduced Hamiltonian}
As mentioned in the introduction, there is in some situations a
subgroup $\Gamma$ of the group of diffeomorphisms of $G$, acting properly
discontinuously on $G$, such that $g(t)$ is invariant under 
$\Gamma$ for all $t$. In such situations, we can consider $g$ to be 
a metric on $M=G/\Gamma$. 
Observe however that the basis $e_{i}'$ need not necessarily
descend to a basis on $M$. Let us
assume $M$ to be compact. We can then consider the reduced
Hamiltonian
\[
H(t)=-[-\theta(t)]^{3}V(t)
\]
where $V(t)=\mathrm{vol}(M,g(t))$. 

\begin{lemma}\label{lemma:reduced}
In terms of the time coordinate $\tau$ defined by (\ref{eq:taudef})
we have
\begin{equation}\label{eq:reduced}
H(\tau)=\theta^{3}(0)V(0)\exp (-3\int_{0}^{\tau}
q d s).
\end{equation}
\end{lemma}
\textit{Proof}. Since $M$ is compact we can
choose a finite partition of unity $\phi_{i}$ $i=1,...,k$
subordinated to coordinate charts. We can then write
\[
V(t)=\sum_{i=1}^{k}\int_{M}\phi_{i}\sqrt{\det g(t)(\partial_{j},
\partial_{m})}  d x^{1}... d x^{3}.
\]
We compute
\[
\frac{ d V}{ d t}=\theta V
\]
so that 
\[
\frac{ d V}{ d\tau}=\theta V\frac{ d t}{ d\tau}=
\theta V\frac{3}{\theta}=
3V.
\]
Considering $H$ as a function of $\tau$, we get (\ref{eq:reduced}) using 
(\ref{eq:raychaudhuri2}). $\Box$

\section{Bianchi I}
Bianchi I solutions of (\ref{eq:whsu})-(\ref{eq:constraint}) are of
the form $(\sigma_{+},\sigma_{-},0,0,0)$ where
$(\sigma_{+},\sigma_{-})$ are independent of time and satisfy
$\sigma_{+}^{2}+\sigma_{-}^{2}=1$. 

\section{Bianchi II}
\begin{prop}\label{prop:b2}
A Bianchi II solution of (\ref{eq:whsu})-(\ref{eq:constraint})
with $\No>0$ and $\Nt=\Nth=0$ satisfies
\begin{equation}\label{eq:nol}
\lim_{\tau\rightarrow \infty}\No=0
\end{equation}
and
\begin{equation}\label{eq:spsml}
\lim_{\tau\rightarrow \infty}(\Sp,\Sm)=(\sigma_{+},\sigma_{-})
\end{equation}
where $(\sigma_{+},\sigma_{-})$ belongs to $\mathcal{K}_{1}$.
\end{prop}
\textit{Proof}. Using the constraint (\ref{eq:constraint}) we deduce
that
\[
\Sp'=\frac{3}{2}\No^{2}(2-\Sp).
\]
We wish to apply the monotonicity principle. There are three 
variables. Let $U$ be defined by $\No>0$, $M$ be defined by
(\ref{eq:constraint}) and $F(\Sp,\Sm,\No)=\Sp$. 
Equation (\ref{eq:nol}) follows, since if $\No$ does not converge
to zero, we can construct an $\omega$-limit point in $U$ due to the 
fact that the variables are contained in a compact set.
Combining this with the constraint we deduce
\[
\lim_{\tau\rightarrow \infty}q=2.
\]
Since $\Sp$ is monotone and bounded, it converges. Since $q\rightarrow 2$
and the $\omega$-limit set is connected, c.f. Lemma \ref{lemma:als},
(\ref{eq:spsml}) holds, but we do not yet know anything
about the limit. Compute
\[
\left(\frac{\Sm}{2-\Sp}\right)'=0.
\]
We get
\begin{equation}\label{eq:b2l}
\frac{\Sm}{2-\Sp}=\frac{\sigma_{-}}{2-\sigma_{+}}
\end{equation}
for arbitrary $(\Sp,\Sm)$ belonging to the solution. Since
$\No'=(q-4\Sp)\No$ and $\No\rightarrow 0$ we have to have
$\sigma_{+}\geq \frac{1}{2}$. If $\sigma_{+}=\frac{1}{2}$ 
then $\sigma_{-}=
\pm\frac{\sqrt{3}}{2}$. The two corresponding lines in the $\Sp\Sm$-plane
obtained by substituting $(\sigma_{+},\sigma_{-})$ into
(\ref{eq:b2l}) do not intersect any points interior to the Kasner
circle. Therefore $\sigma_{+}=\frac{1}{2}$ is not an allowed limit point
and the proposition follows. $\Box$

\section{Bianchi VI${}_{0}$}
\begin{prop}\label{prop:b6}
Consider a Bianchi V$I_{0}$ solution of
(\ref{eq:whsu})-(\ref{eq:constraint}) with $\No=0$, $\Nt>0$ and
$\Nth<0$. Then
\begin{equation}\label{eq:n2n3l}
\lim_{\tau\rightarrow \infty}\Nt=0,\
\lim_{\tau\rightarrow \infty}\Nth=0
\end{equation}
and 
\begin{equation}\label{eq:spsmlim}
\lim_{\tau\rightarrow \infty}\Sm=0,\
\lim_{\tau\rightarrow \infty}\Sp=-1.
\end{equation}
\end{prop}
\textit{Proof}. By the constraint (\ref{eq:constraint}) the variables
belong to a compact set.
Since 
\[
\Sp'=-(2-q)(1+\Sp)
\]
$\Sp$ is strictly monotone as long as $q<2$ which is true for the
entire solution. The monotonicity principle
yields
\begin{equation}\label{eq:qlim}
\lim_{\tau\rightarrow \infty}q=2
\end{equation}
for similar reasons as in the proof of Proposition \ref{prop:b2}.
By the constraint we conclude that (\ref{eq:n2n3l}) holds. 
Since $\Sp$ is monotone, (\ref{eq:qlim}) holds and the $\omega$-limit
set is connected, we have
\[
\lim_{\tau\rightarrow \infty}(\Sp,\Sm)=(\sigma_{+},\sigma_{-})
\]
with $\sigma_{+}^{2}+\sigma_{-}^{2}=1$. Since $\Nt$ and $\Nth$
converge to zero and since they satisfy (\ref{eq:whsu}) we must have
\begin{eqnarray*}
& 2(\sigma_{+}^{2}+\sigma_{-}^{2})+2\sigma_{+}+2\sqrt{3}\sigma_{-}\leq
0\\
& 2(\sigma_{+}^{2}+\sigma_{-}^{2})+2\sigma_{+}-2\sqrt{3}\sigma_{-}\leq
0.
\end{eqnarray*}
Adding these inequalities we obtain $4+4\sigma_{+}\leq 0$.
Equation (\ref{eq:spsmlim}) follows. $\Box$

Let us now record the fact that Bianchi VIII is very different from
Bianchi IX in the direction towards the singularity.

\begin{prop}\label{prop:wierd}
For every $\epsilon>0$, there exists Bianchi VIII vacuum initial data 
$y_{\epsilon}$ and
a real number $T_{\epsilon}>0$, where $y_{\epsilon}$ is closer to the 
Kasner circle 
than $\epsilon$, such that evolving $y_{\epsilon}$ a time $T_{\epsilon}$ 
to the past 
yields a Bianchi VIII point with the property that
\[
|\No\Nt|+|\Nt\Nth|+|\Nth\No|\geq \frac{1}{4}.
\]
\end{prop}
\textit{Remark}. By distance to the Kasner circle we here mean the Euclidean
distance in $\mathbb{R}^{5}$.

\textit{Proof}.
Let $\epsilon>0$ and $x$ be any Bianchi VI${}_{0}$ initial data.
By Proposition \ref{prop:b6}, there is a $T_{\epsilon}>0$ such that 
applying the
flow to $x$, $\Phi(T_{\epsilon},x)$ is closer to the Kasner circle than 
$\epsilon/2$, 
where $\Phi(\tau,x)$ represents the solution to 
(\ref{eq:whsu})-(\ref{eq:constraint}) with initial data $x$ evaluated in
$\tau$. However, there are Bianchi VIII points as close to 
$\Phi(T_{\epsilon},x)$ as
we wish. Given an $\eta>0$, we can thus, by the continuity of the flow,
choose Bianchi VIII initial data $y_{\epsilon}$ with the property that 
$y_{\epsilon}$ is
closer to the Kasner circle than $\epsilon$ and 
$\Phi(-T_{\epsilon},y_{\epsilon})$ is closer to 
$x$ than $\eta$. Choosing $x$ to have $\Sp=\Sm=0$ and $\Nt=-\Nth=
\frac{1}{\sqrt{3}}$ 
and letting $\eta$ be small enough, we get the conclusion of the proposition. 
$\Box$

\section{Bianchi VII${}_{0}$ }\label{section:bsz}
The following proposition is a consequence of  
(\ref{eq:whsu})-(\ref{eq:constraint}).
\begin{prop}\label{prop:b7}
Consider a Bianchi VI$I_{0}$ solution to
(\ref{eq:whsu})-(\ref{eq:constraint}) with $\Nt=\Nth>0$, $\No=0$ and
$\Sm=0$. Then one of the following statements holds
\begin{itemize}
\item $\Sp=1$ for the entire solution and $\Nt=\Nth\rightarrow \infty$
  as $\tau\rightarrow \infty$.
\item $\Sp=-1$ and $\Nt=\Nth$ are constant for the entire solution.
\end{itemize}
\end{prop} 
Given a solution to (\ref{eq:whsu})-(\ref{eq:constraint}),
if $\Nt(\tau)=\Nth(\tau)$ and $\Sm(\tau)=0$ for one $\tau$ then 
the two equalities hold for all $\tau$. Thus the above
proposition together with the one below exhaust the possibilities
for Bianchi VI$\mathrm{I}_{0}$ solutions.

\begin{prop}\label{prop:bsz}
Consider a Bianchi VII${}_{0}$ solution to
(\ref{eq:whsu})-(\ref{eq:constraint}) with $\Nt,\ \Nth>0$ and  $\No=0$
which never satisfies $\Nt=\Nth$ and $\Sm=0$ simultaneously. Then
\[
\lim_{\tau\rightarrow \infty}\Sp=-1,\
\lim_{\tau\rightarrow \infty}\Sm=0
\]
and
\[
\lim_{\tau\rightarrow \infty}(\Nt-\Nth)=0.
\]
\end{prop}
\textit{Remark}. In Proposition \ref{prop:conv}, we also prove that
$\Nt$ and $\Nth$ converge to finite positive values.

\textit{Proof}. \textbf{Assume first that there is an $\omega$-limit
point}. Then the monotonicity principle suffices. Consider
\begin{equation}\label{eq:spr}
\Sp'=-(2-q)(\Sp+1).
\end{equation}
We apply the monotonicity principle with $F(\Sp,\Sm,\Nt,\Nth)=\Sp$,
$M$ defined by (\ref{eq:constraint}) and $U$ defined by
$\Sm^2+(\Nt-\Nth)^2>0$ and $\Nt+\Nth>0$. $F$ is strictly monotone
on a solution in $U\cap M$ since if $\Sp'(\tau)=0$ then 
$[\Nt-\Nth](\tau)=0$ so that $\Sm(\tau)\neq 0$ 
whence $[\Nt-\Nth]'(\tau)\neq 0$, c.f.
(\ref{eq:whsu})-(\ref{eq:constraint}). If $\tau_{k}\rightarrow \infty$
yields an $\omega$-limit point, we must thus have 
$[\Sm^2+(\Nt-\Nth)^2](\tau_{k})\rightarrow 0$ or 
$(\Nt+\Nth)(\tau_{k})\rightarrow 0$ by the monotonicity principle.
However
\[
Z_{-1}=\frac{\frac{4}{3}\Sm^2+(\Nt-\Nth)^{2}}{\Nt\Nth}=
\frac{\frac{4}{3}(1-\Sp^{2})}{\Nt\Nth}
\]
has non-positive derivative, and consequently
$[\Sm^2+(\Nt-\Nth)^2](\tau_{k})\rightarrow 0$ (if it does not 
converge to zero, $Z_{-1}(\tau_{k})\rightarrow \infty$). We conclude that
$\Sp(\tau_{k})\rightarrow -1$. By the monotonicity of $\Sp$ 
the proposition follows.

\textbf{Assume that there is no $\omega$-limit point}.
If there is a sequence $\tau_{k}\rightarrow \infty$ such that 
$\Nt(\tau_{k})$ is bounded, then  there is an $\omega$-limit point
by the constraint (\ref{eq:constraint}). Similarly if there is a 
sequence such that $\Nth$ is bounded. We can thus assume that $\Nt$
and $\Nth$ converge to $\infty$. 

The variable $\Sp$ is still decreasing and bounded. We can thus
assume it to converge to $\sigma_{+}$ with $-1\leq\sigma_{+}<1$. If
$\sigma_{+}=-1$ we are done so assume not. We prove that this
assumption leads to the consequence that  $(2-q)\notin
L^{1}([0,\infty))$. Combining this with (\ref{eq:spr}) we conclude 
that $\Sp\rightarrow -1$ contradicting our assumption.

The intuitive idea is as follows. $\Sp\rightarrow \sigma_{+}$ but
$\Sm$ will oscillate between $\pm (1-\sigma_{+}^{2})^{1/2}$, roughly
speaking. When $\Sm$ is small we get a contribution to the integral
of $2-q$. If $\Sm$ spends most of its time close to 
$\pm (1-\sigma_{+}^{2})^{1/2}$ we would have a problem, but that
turns out not to be the case.

If $\Nt$ and $\Nth$ converge to infinity, we will see that the
solution will become oscillatory. To be more precise,
let $x$ and $y$ be defined by
\begin{eqnarray}
x & = & \frac{\Sm}{(1-\Sp^{2})^{1/2}}\label{eq:xdef1}\\
y & = & \frac{\sqrt{3}}{2}\frac{\Nt-\Nth}{(1-\Sp^{2})^{1/2}}.
\label{eq:ydef1}
\end{eqnarray}
Since $\Sp^{2}(\tau)<1$ for all $\tau$,  $x$ and $y$ are smooth. 
Let
\begin{equation}\label{eq:gdef}
g=3(\Nt+\Nth)+2(1+\Sp)xy.
\end{equation}
Then $x'=-gy$ and $y'=gx$. By the constraint $x^2+y^2=1$ so that
we can choose a $\phi_{0}$ such that $(x(\tau_{0}),y(\tau_{0}))=
(\cos (\phi_{0}),\sin (\phi_{0}))$. Define
\begin{equation}\label{eq:xidef1}
\xi(\tau)=\int_{\tau_{0}}^{\tau}g(s) d s+\phi_{0}.
\end{equation}
Then $x(\tau)=\cos(\xi(\tau))$ and $y(\tau)=\sin (\xi(\tau))$.

As mentioned it suffices to prove that $2-q\notin L^{1}([0,\infty))$
under the assumption that $-1<\sigma_{+}<1$. Consider
\[
 2-q=2(1-\Sp^2-(1-\Sp^2)\cos^2(\xi))=2(1-\Sp^2)\sin^2(\xi)\geq
2(1-\sigma_{+}^2)\sin^2(\xi)
\]
\[
= c\sin^2 (\xi),
\]
where $c>0$ by assumption. Consider a period. Assume $T$ is
great enough that $1<5(\Nt+\Nth)/2\leq g\leq 7(\Nt+\Nth)/2$ for
$\tau\geq T$ and that $\xi(\tau_{2})-\xi(\tau_{1})=2\pi$ 
where $T\leq\tau_{1}<\tau_{2}$. Let us estimate $\Delta\tau=
\tau_{2}-\tau_{1}$. By considering (\ref{eq:xidef1}) we
conclude that $\Delta\tau$ can be chosen arbitrarily small
if $T$ is great enough since $\Nt,\Nth\rightarrow \infty$. Using
(\ref{eq:whsu}) we derive the consequence that for every $\epsilon>0$
\[
1-\epsilon\leq \frac{(\Nt+\Nth)(\tau)}{(\Nt+\Nth)(\tau_{1})}
\leq 1+\epsilon
\]
for every $\tau\in [\tau_{1},\tau_{2}]$ if $T$ is great enough.
Thus
\[
 \Delta\tau=\int_{\tau_{1}}^{\tau_{2}} d\tau=
\int_{\xi(\tau_{1})}^{\xi(\tau_{2})}\frac{1}{g} d\xi\leq
\int_{\xi(\tau_{1})}^{\xi(\tau_{2})}\frac{2}{5(\Nt+\Nth)} d\xi\leq
\frac{8\pi}{5(\Nt+\Nth)(\tau_{1})}
\]
if $T$ is great enough. But then
\[
 \int_{\tau_{1}}^{\tau_{2}}(2-q) d\tau\geq 
\int_{\tau_{1}}^{\tau_{2}}c\sin^{2}(\xi) d\tau=
\int_{\xi(\tau_{1})}^{\xi(\tau_{2})}c\sin^{2}(\xi)\frac{1}{g} d\xi
\]
\[
\geq\int_{\xi(\tau_{1})}^{\xi(\tau_{2})}c\sin^{2}(\xi)
\frac{2}{7(\Nt+\Nth)} d\xi
\geq \frac{c}{7(\Nt+\Nth)(\tau_{1})}
\int_{\xi(\tau_{1})}^{\xi(\tau_{2})}\sin^{2}(\xi) d\xi
\]
\[
= \frac{c\pi}{7(\Nt+\Nth)(\tau_{1})}\geq
\frac{5c}{56}\Delta\tau=c'\Delta\tau
\]
if $T$ is great enough. Here $c'>0$. Let $M>0$ be any number. Consider
an interval $[\tau_{a},\tau_{b}]$ where $\tau_{a}$ is great enough
that the above conditions are met, $(\tau_{b}-\tau_{a})/c'>M$ and
$\xi(\tau_{b})-\xi(\tau_{a})$ is an integer multiple of $2\pi$. Then
\[
\int_{\tau_{a}}^{\tau_{b}}(2-q) d\tau\geq M
\]
and consequently $2-q\notin L^{1}([0,\infty))$. The proposition
follows. $\Box$ 

It will be of interest to know that $\Nt$ is bounded away from
zero. 
\begin{lemma}\label{lemma:bsz}
Consider a Bianchi VII${}_{0}$ solution to 
(\ref{eq:whsu})-(\ref{eq:constraint}). Then
\[
\Nt\Nth\geq c>0
\]
on $[0,\infty)$, where $c$ is a positive constant.
\end{lemma}
\textit{Proof}. Assume there is a time sequence $\tau_{k}
\rightarrow \infty$ such that $(\Nt\Nth)(\tau_{k})$ converges
to zero. Since $\Nt-\Nth\rightarrow 0$, we conclude that
the solution evaluated at $\tau_{k}$ converges to the Kasner
circle. Since the flow is continuous, and Bianchi VII${}_{0}$ belongs
to the boundary of Bianchi IX, we can construct a sequence of 
Bianchi IX initial data $x_{k}$ converging to the Kasner circle,
such that $\Phi(-\tau_{k},x_{k})$ converges to a Bianchi VII${}_{0}$
point with $\Nt\Nth>0$. This is impossible by Theorem 15.1
of \cite{jag2}. $\Box$

As was noted in the proof of Proposition \ref{prop:bsz}, the
objects $x$ and $y$ defined in (\ref{eq:xdef1}) and (\ref{eq:ydef1})
can be written as 
\[
x(\tau)=\cos \xi(\tau),\ \ y(\tau)=\sin \xi(\tau)
\]
where $\xi$ is defined in (\ref{eq:xidef1}). Since $\Nt+\Nth$ is
bounded away from zero to the future by Lemma \ref{lemma:bsz}
and $1+\Sp$ converges to zero, we can assume that 
\begin{equation}\label{eq:gbound}
g(\tau)\geq 2(\Nt+\Nth)(\tau)\geq c>0\ \ \forall \tau\geq \tau_{0}
\end{equation}
for some positive constant $c$, where 
$g$ is defined by (\ref{eq:gdef}). In other words, the expressions
$x$ and $y$ will carry out an infinite number of oscillations as
time progresses. It is natural to average over a period of this 
oscillation in order to analyze the asymptotics, and we need to 
know how certain expressions vary over a period to be able to
do this. 

In the rest of this section we will implicitly assume that all
Bianchi VII${}_{0}$ solutions satisfy $\Sm^2+(\Nt-\Nth)^2>0$.

\begin{lemma}
Consider a Bianchi VII${}_{0}$ solution to 
(\ref{eq:whsu})-(\ref{eq:constraint}) with $\Nt,\Nth> 0$,
and let $\tau_{0}$ be such that (\ref{eq:gbound}) is fulfilled.
Let $\tau_{0}\leq \tau_{a}< \tau_{b}$, and assume
\[
\xi(\tau_{b})-\xi(\tau_{a})=\pi.
\]
Then there is a $T$ such that 
\begin{equation}\label{eq:spvar3}
|1-\frac{(1+\Sp)(\tau_{1})}
{(1+\Sp)(\tau_{2})}|\leq C
\frac{(1+\Sp)(\tau_{b})}{(\Nt+\Nth)(\tau_{\max})}
\end{equation}
and
\begin{equation}\label{eq:nvar3}
|1-\frac{(\Nt+\Nth)(\tau_{1})}
{(\Nt+\Nth)(\tau_{2})}|\leq C
\frac{(1+\Sp)(\tau_{b})}{(\Nt+\Nth)(\tau_{\max})}
\end{equation}
if $\tau_{a}\geq T$, where $\tau_{\max}$ yields the maximum value
of $\Nt+\Nth$ in $[\tau_{a},\tau_{b}]$ and 
$\tau_{1},\tau_{2}$ are arbitrary members of $[\tau_{a},\tau_{b}]$. 
The constant $C$ only depends on 
the constant $c$ appearing in (\ref{eq:gbound}).
\end{lemma}
\textit{Proof}. Observe that
\[
\pi=\int_{\tau_{a}}^{\tau_{b}}g(s) d s\geq c(\tau_{b}-\tau_{a}),
\]
so that
\begin{equation}\label{eq:dtbound}
\tau_{b}-\tau_{a}\leq \frac{\pi}{c}.
\end{equation}
Consider
\[
(\Nt+\Nth)'=(q+2\Sp)(\Nt+\Nth)+2\sqrt{3}\Sm(\Nt-\Nth).
\]
Estimate
\[
 |q+2\Sp|\leq 2|\Sp(1+\Sp)|+2\Sm^2\leq 2(1+\Sp)+2(1-\Sp^2)\leq
6(1+\Sp),
\]
and
\[
 |2\sqrt{3}\Sm (\Nt-\Nth)|\leq 2[\Sm^2+\frac{3}{4}(\Nt-\Nth)^2]
\leq 2(1-\Sp^2)\leq 4(1+\Sp).
\]
Since $\Nt+\Nth$ is bounded below by a positive constant for 
$\tau\geq\tau_{0}$, we conclude that
\[
|(\Nt+\Nth)'|\leq C_{1} (1+\Sp)(\Nt+\Nth)
\]
for some positive constant $C_{1}$. Letting $\tau_{1},\tau_{2}
\in [\tau_{a},\tau_{b}]$, we get
\begin{equation}\label{eq:whatnot}
\ln\alpha=\ln\left(\frac{(\Nt+\Nth)(\tau_{1})}
{(\Nt+\Nth)(\tau_{2})}\right)\leq
C_{1}\int_{\tau_{a}}^{\tau_{b}}(1+\Sp) d s=\beta,
\end{equation}
where we have introduced the quantities $\alpha$ and $\beta$ in
order to make the following arguments easier to follow. If
$\alpha\geq 1$, then
\begin{equation}\label{eq:whatnot2}
0\leq \alpha-1\leq  e^{\beta}-1=\sum_{k=1}^{\infty}\frac{\beta^k}{k!}\leq
\beta  e^{\beta}.
\end{equation}
If $\alpha\leq 1$, then (\ref{eq:whatnot2}) also holds if we replace 
$\alpha$ with $1/\alpha$, since $\tau_{1}$ and $\tau_{2}$ in 
(\ref{eq:whatnot}) are arbitrary. Multiplying this inequality with 
$\alpha$ then yields the conclusion
\[
|1-\alpha|\leq \beta  e^{\beta}
\]
without any conditions on $\alpha$. 
Since we have the bound (\ref{eq:dtbound}) and $0\leq 1+\Sp\leq 2$,
we conclude that $ e^{\beta}$ is bounded by a constant, so that 
\begin{equation}\label{eq:nvar1}
|1-\frac{(\Nt+\Nth)(\tau_{1})}
{(\Nt+\Nth)(\tau_{2})}|\leq C_{2}
\int_{\tau_{a}}^{\tau_{b}}(1+\Sp) d s.
\end{equation}
Observe that this inequality proves that the left hand side can be 
chosen to be arbitrarily small by 
demanding $\tau_{a}$ to be big enough. Let us estimate the integral
on the right hand side. Assuming $\tau_{a}$ to be great enough
that the right hand side of (\ref{eq:nvar1}) is less than $\frac{1}{2}$,
and using (\ref{eq:gbound}) and the fact that $\Sp$ is monotone,
we conclude that
\[
 \int_{\tau_{a}}^{\tau_{b}}(1+\Sp) d s\leq
(1+\Sp)(\tau_{a})\int_{\xi(\tau_{a})}^{\xi(\tau_{b})}\frac{1}{g} d\xi\leq
(1+\Sp)(\tau_{a})\int_{\xi(\tau_{a})}^{\xi(\tau_{b})}
\frac{1}{2(\Nt+\Nth)} d\xi
\]
\[
\leq
\frac{3\pi}{4}\frac{(1+\Sp)(\tau_{a})}{(\Nt+\Nth)(\tau_{\max})}
\]
where $\tau_{\max}$ corresponds to the maximum value of $\Nt+\Nth$
in $[\tau_{1},\tau_{2}]$. Combining this with (\ref{eq:nvar1})
we get
\begin{equation}\label{eq:nvar2}
|1-\frac{(\Nt+\Nth)(\tau_{1})}
{(\Nt+\Nth)(\tau_{2})}|\leq C_{3}
\frac{(1+\Sp)(\tau_{a})}{(\Nt+\Nth)(\tau_{\max})}.
\end{equation}
Consider now 
\begin{equation}\label{eq:sppr1}
(1+\Sp)'=-(2-q)(1+\Sp)
\end{equation}
Since 
\[
0\leq 2-q\leq 4(1+\Sp),
\]
we conclude that 
\begin{equation}\label{eq:spvar1}
|1-\frac{(1+\Sp)(\tau_{1})}
{(1+\Sp)(\tau_{2})}|\leq C_{3}
\frac{(1+\Sp)(\tau_{a})}{(\Nt+\Nth)(\tau_{\max})}
\end{equation}
by an argument similar to that proving (\ref{eq:nvar2}).
By assuming $\tau_{a}$ to be big enough (\ref{eq:spvar1})
leads to the conclusion that we, at the cost of increasing the 
constants, can replace 
$(1+\Sp)(\tau_{a})$ in the right hand side of (\ref{eq:nvar2})
and (\ref{eq:spvar1}) with the same expression evaluated in
$\tau_{b}$, yielding the statement of the lemma. $\Box$

The following lemma constitutes the main observation.

\begin{lemma}\label{lemma:oscvar}
Consider a Bianchi VII${}_{0}$ solution to 
(\ref{eq:whsu})-(\ref{eq:constraint}) with $\Nt,\Nth> 0$,
and let $\tau_{0}$ be such that (\ref{eq:gbound}) is fulfilled.
Let $\tau_{0}\leq \tau_{a}< \tau_{b}$, and assume
\[
\xi(\tau_{b})-\xi(\tau_{a})=\pi.
\]
Then there is a $T$ such that 
\begin{equation}\label{eq:oscvar}
|\int_{\tau_{a}}^{\tau_{b}}[\Sm^2-(1+\Sp)] d s|\leq
C\int_{\tau_{a}}^{\tau_{b}}(1+\Sp)^2 d s
\end{equation}
if $\tau_{a}\geq T$, for some constant $C$ depending only on the
constant $c$ appearing in (\ref{eq:gbound}).
\end{lemma}
\textit{Proof}. Observe first that
\[
\Sm^2=(1-\Sp^2)x^2=(1-\Sp)(1+\Sp)x^{2}=2(1+\Sp)x^2-(1+\Sp)^{2}x^2.
\]
In consequence, the integrand of interest is 
\[
2(1+\Sp)x^2-(1+\Sp)=(1+\Sp)\cos \eta
\]
where $\eta(\tau)=2\xi(\tau)$. Let $\eta_{a}=\eta(\tau_{a})$ and
$\eta_{b}=\eta(\tau_{b})=\eta_{a}+2\pi$ and observe that
\[
\int_{\tau_{a}}^{\tau_{b}}(1+\Sp)\cos\eta  d \tau=
\int_{\eta_{a}}^{\eta_{b}}\frac{1+\Sp}{2g}\cos\eta d \eta
\]
\[
=
\int_{\eta_{a}}^{\eta_{a}+\pi}\left(
\frac{1+\Sp(\eta)}{2g(\eta)}-\frac{1+\Sp(\eta+\pi)}{2g(\eta+\pi)}
\right)\cos\eta d \eta
\]
where we in order to obtain the last equality divided the integral
into one part from $\eta_{a}$ to $\eta_{a}+\pi$ and one from
$\eta_{a}+\pi$ to $\eta_{a}+2\pi$, and then made the substitution
$\eta\rightarrow \eta-\pi$ in the second integral. We now wish to
prove that the absolute value of the integrand in the integral from
$\eta_{a}$ to $\eta_{a}+\pi$ can be bounded by
\[
C_{1}\frac{[1+\Sp(\eta)]^2}{2g(\eta)}+C_{2}
\frac{[1+\Sp(\eta+\pi)]^2}{2g(\eta+\pi)},
\]
since this would prove the assertion. Let us first rewrite 
\[
\frac{1+\Sp(\eta)}{2g(\eta)}-\frac{1+\Sp(\eta+\pi)}{2g(\eta+\pi)}=
\frac{1+\Sp(\eta)}{2g(\eta)}\left(
1-\frac{g(\eta)}{g(\eta+\pi)}\right)-
\frac{\Sp(\eta+\pi)-\Sp(\eta)}{2g(\eta+\pi)}.
\]
Using (\ref{eq:spvar3}) and the fact that $\Nt+\Nth$ is bounded from
below by a positive constant, we note that the second term can be 
estimated in the desired way; multiply (\ref{eq:spvar3}) by
$1+\Sp(\tau_{2})$ in order to get an estimate of $\Sp(\tau_{1})-
\Sp(\tau_{2})$. If we had 
\[
1-\frac{(\Nt+\Nth)(\eta)}{(\Nt+\Nth)(\eta+\pi)}
\]
instead of 
\[
1-\frac{g(\eta)}{g(\eta+\pi)}
\]
in the first term, then we could estimate the first term in the desired
way as well. However, using (\ref{eq:spvar3}), (\ref{eq:nvar3}) and the 
fact that $g$ and $\Nt+\Nth$ can be bounded from below by a positive
constant, one can show that 
\[
\frac{(\Nt+\Nth)(\eta)}{(\Nt+\Nth)(\eta+\pi)}
-\frac{g(\eta)}{g(\eta+\pi)}
\]
is small enough to yield the desired bound. $\Box$
\begin{lemma}
Consider a Bianchi VII${}_{0}$ solution to 
(\ref{eq:whsu})-(\ref{eq:constraint}) with $\Nt,\Nth> 0$,
and let $\tau_{0}$ be such that (\ref{eq:gbound}) is fulfilled.
Then
\begin{equation}\label{eq:intconv1}
\lim_{\tau\rightarrow\infty}\frac{\int_{\tau_{0}}^{\tau}\Sm^2 d s}
{\int_{\tau_{0}}^{\tau}(1+\Sp) d s}=1
\end{equation}
and
\begin{equation}\label{eq:intconv2}
\lim_{\tau\rightarrow\infty}\frac{\int_{\tau_{0}}^{\tau}(2-q) d s}
{\int_{\tau_{0}}^{\tau}2(1+\Sp) d s}=1.
\end{equation}
\end{lemma}
\textit{Proof}. Observe first that since (\ref{eq:sppr1}) is fulfilled
and $1+\Sp\rightarrow 0$ as $\tau\rightarrow \infty$,
$2-q\notin L^{1}([\tau_{0},\infty))$. Moreover,
\[
2-q\leq 2(1-\Sp^2)\leq 4(1+\Sp),
\]
so that $1+\Sp\notin L^{1}([\tau_{0},\infty))$.
In order to prove (\ref{eq:intconv1}), let us
prove that
\[
\lim_{\tau\rightarrow\infty}
\frac{\int_{\tau_{0}}^{\tau}[\Sm^2-(1+\Sp)] d s}
{\int_{\tau_{0}}^{\tau}(1+\Sp) d s}=0.
\]
Let $\epsilon>0$ and let $T$ be such that Lemma \ref{lemma:oscvar} 
can be applied and such that $1+\Sp(\tau)\leq \epsilon/(2C)$ for
$\tau\geq T$, where $C$ is the constant appearing in (\ref{eq:oscvar}).
Given $\tau>T$, let $\tau_{1}$ be the greatest number smaller than
$\tau$ such that $\xi(\tau_{1})-\xi(T)$ is an integer multiple of
$\pi$. We have
\[
 |\int_{\tau_{0}}^{\tau}[\Sm^2-(1+\Sp)] d s|\leq
\int_{\tau_{0}}^{T}|\Sm^2-(1+\Sp)| d s+
|\int_{T}^{\tau_{1}}[\Sm^2-(1+\Sp)] d s|
\]
\[
+\int_{\tau_{1}}^{\tau}|\Sm^2-(1+\Sp)| d s.
\]
We should divide this expression with $\int_{\tau_{0}}^{\tau}
(1+\Sp) d s$, and then evaluate the limit as $\tau\rightarrow
\infty$. After having carried out this division, the first and the
third terms will yield something that can be assumed to be 
arbitrarily small by assuming $\tau$ to be big enough. Using
(\ref{eq:oscvar}), the middle term can be bounded by
\[
 |\int_{T}^{\tau_{1}}[\Sm^2-(1+\Sp)] d s|\leq
C\int_{T}^{\tau_{1}}(1+\Sp)^2 d s\leq
C[1+\Sp(T)]\int_{T}^{\tau_{1}}(1+\Sp) d s
\]
\[
\leq
\frac{\epsilon}{2}\int_{T}^{\tau_{1}}(1+\Sp) d s.
\]
After having carried out the division, the corresponding contribution
can consequently be assumed to be less than or equal to $\epsilon/2$. 
Equation (\ref{eq:intconv1}) follows. In order to prove 
(\ref{eq:intconv2}), observe that
\[
(2-q)-2(1+\Sp)=
2(1+\Sp)-2\Sm^2-2(1+\Sp)^2.
\]
In consequence the proof of (\ref{eq:intconv2}) is similar
to the proof of (\ref{eq:intconv1}). $\Box$

\begin{lemma}\label{lemma:conv}
Consider a Bianchi VII${}_{0}$ solution to 
(\ref{eq:whsu})-(\ref{eq:constraint}) with $\Nt,\Nth> 0$,
and let $\tau_{0}$ be such that (\ref{eq:gbound}) is fulfilled.
Then, if $p>1$,
\[
1+\Sp\in L^{p}([\tau_{0},\infty)).
\]
\end{lemma}
\textit{Proof}. By (\ref{eq:intconv2}) we conclude that
for each $\alpha_{1}<2<\alpha_{2}$, there is a $T$ such that
$\tau\geq T$ implies
\[
\alpha_{1}<
\frac{\int_{\tau_{0}}^{\tau}(2-q) d s}
{\int_{\tau_{0}}^{\tau}(1+\Sp) d s}<\alpha_{2}.
\]
Introduce the notation
\[
\alpha=1+\Sp(\tau_{0})
\]
and
\[
h(\tau)=\int_{\tau_{0}}^{\tau}[1+\Sp(s)] d s.
\]
By (\ref{eq:sppr1}) we conclude that
\begin{equation}\label{eq:auxineq}
\alpha\exp[-\alpha_{2}h(\tau)]\leq
h'(\tau)\leq
\alpha\exp[-\alpha_{1}h(\tau)]
\end{equation}
for all $\tau\geq T$. Integrating the leftmost inequality, we get
\[
\frac{1}{\alpha_{2}}
\ln [\exp[\alpha_{2}h(T)]+\alpha\alpha_{2}(\tau-T)] \leq h(\tau)
\]
for all $\tau\geq T$. Inserting this in the rightmost inequality of
(\ref{eq:auxineq}), we conclude that
\[
h'(\tau)\leq\alpha 
[\exp[\alpha_{2}h(T)]+\alpha\alpha_{2}(\tau-T)]^{-
\frac{\alpha_{1}}{\alpha_{2}}}.
\]
Since $h'=1+\Sp$, and since the quotient $\alpha_{1}/\alpha_{2}$
can be chosen to be arbitrarily close to one, the lemma follows.
$\Box$

\begin{prop}\label{prop:conv}
Consider a Bianchi VII${}_{0}$ solution to 
(\ref{eq:whsu})-(\ref{eq:constraint}) with $\Nt,\Nth> 0$.
Then
\[
\lim_{\tau\rightarrow\infty}\Nt(\tau)=
\lim_{\tau\rightarrow\infty}\Nth(\tau)=n_{0}
\]
where $0<n_{0}<\infty$.
\end{prop}
\textit{Proof}. It is enough to prove that $\Nt\Nth$ converges to a
positive number. 
The reason is that this would prove that all the variables are contained 
in a compact set to the future, so that
\[
\Nt=(\Nt\Nth)^{1/2}+\Nt^{1/2}\frac{\Nt-\Nth}{\Nt^{1/2}+\Nth^{1/2}}
\]
converges, since $\Nt\Nth$ converges, $\Nt^{1/2}+\Nth^{1/2}$
is bounded from below by a positive constant, $\Nt^{1/2}$ is bounded 
from above, and
$\Nt-\Nth\rightarrow 0$ by Proposition \ref{prop:bsz}. Since
$\Nt-\Nth\rightarrow 0$, the limit for $\Nth$ follows.
Since
\[
(\Nt\Nth)'=(2q+4\Sp)\Nt\Nth,
\]
the essential point is to prove that 
\[
\int_{\tau_{0}}^{\tau}(q/2+\Sp) d s
\]
converges as $\tau\rightarrow \infty$. Observe now that
\[
q/2+\Sp=\Sp(1+\Sp)+\Sm^2=(1+\Sp)^2-(1+\Sp)+\Sm^2.
\]
Since $1+\Sp\in L^{2}([\tau_{0},\infty))$ by Lemma \ref{lemma:conv}, 
we need only consider
\[
\psi(\tau)=\int_{\tau_{0}}^{\tau}[\Sm^2-(1+\Sp)] d s.
\]
Let $T$ be big enough that (\ref{eq:oscvar}) is applicable.
If $\tau_{2}\geq\tau_{1}\geq T$ and $\tau_{3}$ is the largest
number smaller than $\tau_{2}$ such that 
\[
\xi(\tau_{3})-\xi(\tau_{1})
\]
is an integer multiple of $\pi$, then
\[
\psi(\tau_{2})-\psi(\tau_{1})=
\int_{\tau_{1}}^{\tau_{3}}[\Sm^2-(1+\Sp)] d s+
\int_{\tau_{3}}^{\tau_{2}}[\Sm^2-(1+\Sp)] d s.
\]
The second integral converges to zero as $\tau_{1}$ and $\tau_{2}$ go 
to infinity, since the integrand goes to zero and the length of the
interval is bounded (\ref{eq:dtbound}). By (\ref{eq:oscvar}) we 
conclude that
\[
|\int_{\tau_{1}}^{\tau_{3}}[\Sm^2-(1+\Sp)] d s|\leq
C\int_{\tau_{1}}^{\tau_{3}}(1+\Sp)^2 d s.
\]
Since $1+\Sp\in L^{2}([\tau_{0},\infty))$, this expression converges
to zero as $\tau_{1}$ goes to infinity. The lemma follows. $\Box$

\section{Bianchi VIII}
We assume that $\Nt,\ \Nth>0$ and $\No<0$. Let us begin by giving
the asymptotic behaviour of the $N_{i}$. 
\begin{lemma}\label{lemma:n123l} Consider a Bianchi VIII solution to
(\ref{eq:whsu})-(\ref{eq:constraint}) with $\Nt,\ \Nth>0$ and $\No<0$.
Then,
\[
\lim_{\tau\rightarrow \infty}\No (\tau)=0,\ \
\lim_{\tau\rightarrow \infty}\Nt (\tau)=
\lim_{\tau\rightarrow \infty}\Nth (\tau)=\infty.
\]
\end{lemma}
\textit{Proof}. Reformulate the constraint (\ref{eq:constraint})
as
\[
\Sp^2+\Sm^2+\frac{3}{4}[\No^2+(\Nt-\Nth)^2-2\No(\Nt+\Nth)]=1.
\]
Observe that all terms appearing on the left hand side are non-negative
so that the absolute values of, for instance, $\No$ and $\Nt-\Nth$ 
are bounded by numerical constants.
Assume there is a sequence $\tau_{k}\rightarrow \infty$
such that $\Nt(\tau_{k})\leq C<\infty$. Then the
$N_{i}(\tau_{k})$ are uniformly bounded. By the constraint
$\Sp$ and $\Sm$ are contained in a compact set. Thus we can choose
a convergent subsequence so that we have an $\omega$-limit point
$(\sigma_{+},\sigma_{-},n_{1},n_{2},n_{3})$. We apply the monotonicity
principle with $U$ defined by $N_{i}\neq 0$, $i=1,2,3$, $M$ defined
by (\ref{eq:constraint}) and $F(\Sp,\Sm,\No,\Nt,\Nth)=\No\Nt\Nth$.
$F$, evaluated on a solution contained in $U\cap M$, is strictly
monotone since $(\No\Nt\Nth)'=3q\No\Nt\Nth$. If $q(\tau)=0$ then
either $\Sm'(\tau)\neq 0$ or $\Sp'(\tau)\neq 0$ by 
(\ref{eq:whsu})-(\ref{eq:constraint}). By the monotonicity 
principle one of the $n_{i}$ must be zero contradicting the growth
of $|\No\Nt\Nth|$.
We conclude that $\Nt\rightarrow \infty$, but then  $\Nth\rightarrow
\infty$ and $\No\rightarrow 0$ since $\No(\Nt+\Nth)$ and $\Nt-\Nth$
are bounded by the constraint (\ref{eq:constraint}). $\Box$

\subsection{The NUT case}
Consider the special case $\Nt=\Nth$ and $\Sm=0$.
\begin{prop}
A Bianchi VIII solution with $\No<0$, satisfying $\Nt=\Nth>0$ and $\Sm=0$
has the following asymptotic behaviour
\[
 \lim_{\tau\rightarrow \infty}\Sp(\tau)=\frac{1}{2},\
\lim_{\tau\rightarrow \infty}\No(\tau)=0,\
\lim_{\tau\rightarrow \infty}\Nt(\tau)=\infty\ \mathrm{and}\
\lim_{\tau\rightarrow \infty}(\No\Nt)(\tau)=-\frac{1}{4}.
\]
\end{prop}
\textit{Proof}. The constraint implies
\[
\Sp'=(1-\Sp^{2})(1-2\Sp)+\frac{9}{4}\No^{2}.
\]
Observe that we cannot have $\Sp\rightarrow 1$ since $\No\Nt$
diverges to infinity there, c.f. (\ref{eq:whsu}). Also, 
$\Sp\rightarrow -1$ is an impossibility since $\Sp'$ is
always positive when $\Sp$ is close to $-1$.
Since $\No\rightarrow 0$ by Lemma \ref{lemma:n123l} and 
$-1<\Sp(\tau)<1\ \forall \tau$ by the constraint (\ref{eq:constraint})
$\Sp\rightarrow \frac{1}{2}$ follows. The last limit follows from 
the first and the constraint. $\Box$

\subsection{Oscillatory behaviour}
The behaviour of a Bianchi VIII solution as $\tau\rightarrow \infty$
is in some sense oscillatory. The quantities that oscillate are
$\Sm$ and $\Nt-\Nth$. In order to analyze the asymptotics, we 
quantify this behaviour. Let 
\begin{equation}\label{eq:txtydef}
\tx=\Sm,\ \ty=\frac{\sqrt{3}}{2}(\Nt-\Nth).
\end{equation}
We have
\[
\tx'=-3(\Nt+\Nth)\ty+\epsilon_{x},\
\epsilon_{x}=-(2-q)\Sm+3\No\ty.
\]
Furthermore,
\[
\ty'=3(\Nt+\Nth)\tx+\epsilon_{y},\
\epsilon_{y}=(q+2\Sp)\ty.
\]
Observe that by the constraint (\ref{eq:constraint}) $|\tx|,
|\ty|,|\epsilon_{x}|$ and $|\epsilon_{y}|$ are bounded by
numerical constants whereas $\Nt+\Nth\rightarrow \infty$
by Lemma \ref{lemma:n123l}. Let $g=3(\Nt+\Nth)$,
\[
A=\left( \begin{array}{rr}
                0 & -g \\
                g & 0
            \end{array}
     \right)
\]
$\tbx=(\tx, \ty)^{t} $ and
$\epsilon=(\epsilon_{x}, \epsilon_{y})^{t}$ so that
$\tbx'=A\tbx+\epsilon$.
Let 
\begin{equation}\label{eq:xidef}
\xi(\tau)=\int_{\tau_{0}}^{\tau}g(s) d s+\phi_{0}
\end{equation}
for some $\phi_{0}$ and
$x_{1}(\tau)=\cos(\xi(\tau)),\
y_{1}(\tau)=\sin(\xi(\tau)).$
Then, if $\mathbf{x}_{1}=(x_{1}, y_{1})^{t}$,
$\mathbf{x}_{1}'=A\mathbf{x}_{1}$. Define
\[
\Phi=\left( \begin{array}{rr}
                x_{1} & y_{1} \\
                -y_{1} & x_{1}
            \end{array}
     \right).
\]
Then $\Phi'=-A\Phi$ and $[A,\Phi]=0$. Let 
\begin{equation}\label{eq:rdef}
r(\tau)=(\tilde{x}^{2}(\tau)+\tilde{y}^{2}(\tau))^{1/2} 
\end{equation}
and
\begin{equation}\label{eq:xdef}
\mathbf{x}(\tau)=(x(\tau),y(\tau))^{t}=(r(\tau_{0})\cos[\xi(\tau)],
r(\tau_{0})\sin[\xi(\tau)])^{t}
\end{equation}
where  $\phi_{0}$ has been chosen so that $\mathbf{x}(\tau_{0})=
\tilde{\mathbf{x}}(\tau_{0})$. Since
$[\Phi (\mathbf{x}-\tilde{\mathbf{x}})]'=-\Phi\epsilon$
and $\Phi(\tau)\in SO(2)\ \forall \tau$ we have
\[
\|\tilde{\mathbf{x}}(\tau)-\mathbf{x}(\tau)\|\leq
|\int_{\tau_{0}}^{\tau}\|\epsilon(s)\| d s|. 
\]
By the constraint (\ref{eq:constraint}), we have 
$\|\epsilon\|\leq 9$, and thus
\begin{equation}\label{eq:approximation}
\|\tilde{\mathbf{x}}(\tau)-\mathbf{x}(\tau)\|\leq 9 |\tau-\tau_{0}|.
\end{equation}
The next lemma collects some technical estimates. We present them
here in order not to interrupt the flow of later proofs.

\begin{lemma}\label{lemma:tnest}
Consider a Bianchi VIII solution of
(\ref{eq:whsu})-(\ref{eq:constraint}) and let $C$ be a constant. Then
there is a $T$ depending on $C$ and the initial values such that if 
$[\tau_{1},\tau_{2}]$ is a time interval with $\tau_{1}\geq T$
and  
\[
|\tau_{2}-\tau_{1}|\leq \frac{C}{(\Nt+\Nth)(\tau_{3})}
\]
for some $\tau_{3}\in [\tau_{1},\tau_{2}]$, then
\begin{eqnarray}
|\Sp(t_{1})-\Sp(t_{2})|  \leq  \frac{C_{1}}{(\Nt+\Nth)(t_{3})}
\label{eq:spvar}\\
|[\No(\Nt+\Nth)](t_{1})-[\No(\Nt+\Nth)](t_{2})| 
 \leq  \frac{C_{2}}{(\Nt+\Nth)(t_{3})}\label{eq:nvar}\\
|1-\frac{(\Nt+\Nth)(t_{1})}{(\Nt+\Nth)(t_{2})}|  \leq 
\frac{C_{3}}{(\Nt+\Nth)(t_{3})}\label{eq:n23var}
\end{eqnarray}
for arbitrary $t_{1},t_{2},t_{3}\in [\tau_{1},\tau_{2}]$ where 
$C_{1}$, $C_{2}$ and $C_{3}$ are  constants only depending on $C$.
\end{lemma}
\textit{Proof}. By Lemma \ref{lemma:n123l} we can assume $T$ to be 
great enough that $\Delta\tau=\tau_{2}-\tau_{1}\leq 1$. The inequality
\[
|\frac{(\Nt+\Nth)'}{\Nt+\Nth}|\leq 8
\]
follows from (\ref{eq:whsu})-(\ref{eq:constraint}). Thus
\[
|\frac{(\Nt+\Nth)(t_{1})}{(\Nt+\Nth)(t_{2})}|\leq
  e^{8(\tau_{2}-\tau_{1})}
\]
so that if $(\Nt+\Nth)(t_{1})\geq (\Nt+\Nth)(t_{2})$,
\begin{equation}\label{eq:n23varp}
|1-\frac{(\Nt+\Nth)(t_{1})}{(\Nt+\Nth)(t_{2})}|\leq
| e^{8\Delta\tau}-1|\leq 8 e^{8}\Delta\tau
\end{equation}
since $\Delta\tau\leq 1$. Multiplying this inequality by
$(\Nt+\Nth)(t_{2})/(\Nt+\Nth)(t_{1})\leq 1$ we conclude that the
assumption  $(\Nt+\Nth)(t_{1})\geq (\Nt+\Nth)(t_{2})$ is
not essential. There is thus 
a constant $c<\infty$ such that
\[
\frac{(\Nt+\Nth)(t_{1})}{(\Nt+\Nth)(t_{2})}\leq c
\]
for any $t_{1},t_{2}\in [\tau_{1},\tau_{2}]$. Consequently
\begin{equation}\label{eq:detest}
\Delta\tau\leq \frac{C}{(\Nt+\Nth)(\tau_{3})}\leq
\frac{cC}{(\Nt+\Nth)(t_{3})}.
\end{equation}
By (\ref{eq:whsu})-(\ref{eq:constraint}) 
$|\Sp'|$ and $|[\No(\Nt+\Nth)]'|$ are bounded by  
constants independent of Bianchi VIII solution. Equations
(\ref{eq:spvar}) and (\ref{eq:nvar}) follow from (\ref{eq:detest}).
Equation (\ref{eq:n23var}) follows from (\ref{eq:n23varp}) and
(\ref{eq:detest}).
$\Box$

Let us give the intuitive idea behind the next lemma.
How the variables vary during a period is not so interesting, what is
interesting is to compute the change over a period. In that way
the continuous time evolution is replaced with a discrete evolution.
To make the time step independent of approximation we wish
to see what happens from $\tilde{x}=0$ to the next time $\tilde{x}=0$,
but then we need to know that we have such zeros of $\tilde{x}$. In
order to be able to prove this we assume that $r$ (\ref{eq:rdef}) 
is not too small.
The assumption is not too restrictive; our ultimate goal is to prove that 
$r\rightarrow 0$ and that a general Bianchi VIII solution in
some sense converges to a NUT solution. If $r$ is big our iteration
is well defined and we will use it to prove that $r$ decreases and
if $r$ is already small we do not need it.
\begin{lemma}\label{lemma:zeros}
Consider a Bianchi VIII solution of
(\ref{eq:whsu})-(\ref{eq:constraint}).
There is a $T$ depending on the initial values such that if
$\tau_{0}\geq T$ and $r(\tau_{0})\geq
[(\Nt+\Nth)(\tau_{0})]^{-1/2}$, then $\tilde{x}$ has at least four zeros 
$\tau_{a}<\tau_{b}\leq \tau_{0}\leq \tau_{c}<\tau_{d}$ in
\[
[\tau_{1},\tau_{2}]=[\tau_{0}-\frac{\pi}{(\Nt+\Nth)(\tau_{0})},
\tau_{0}+\frac{\pi}{(\Nt+\Nth)(\tau_{0})}]
\]
such that $\tilde{x}$ is non zero in $(\tau_{a},\tau_{b})$,
$(\tau_{b},\tau_{c})$ and $(\tau_{c},\tau_{d})$.
Furthermore, for any two consecutive zeros, for example $\tau_{a}$ and 
$\tau_{b}$, we have
\begin{equation}\label{eq:xivar}
|\xi(\tau_{b})-\xi(\tau_{a})-\pi|\leq
 \frac{C}{[(\Nt+\Nth)(\tau_{a})]^{1/2}}
\end{equation}
where $C$ is a numerical constant.
\end{lemma}
\textit{Proof}. Let $T$ be great enough that if $\tau\in
[\tau_{1},\tau_{2}]$ then 
\begin{equation}\label{eq:nest}
3(\Nt+\Nth)(\tau)\geq
5(\Nt+\Nth)(\tau_{0})/2\geq 5(\Nt+\Nth)(\tau)/4.
\end{equation}
This is possible by Lemma \ref{lemma:tnest} and Lemma
\ref{lemma:n123l}. Let $\mathbf{x}$ be as in (\ref{eq:xdef}) where
$\phi_{0}\in [0,2\pi]$ in (\ref{eq:xidef}) has been chosen so that 
$\mathbf{x}(\tau_{0})=\tilde{\mathbf{x}}(\tau_{0})$. 
If $\tau\in [\tau_{1},\tau_{2}]$ then
\[
|\xi(\tau)-\xi(\tau_{0})|=|\int_{\tau_{0}}^{\tau}g(s) d s|\geq
\frac{5}{2}(\Nt+\Nth)(\tau_{0})|\tau-\tau_{0}|.
\]
Since $\xi$ is a monotonically growing function we conclude
\[
[0,2\pi]\subseteq
[\phi_{0}-\frac{5}{2}\pi,\phi_{0}+\frac{5}{2}\pi]\subseteq 
[\xi(\tau_{1}),\xi(\tau_{2})].
\]
We pick out the zeros in the interval $[0,2\pi]$.
Let $\xi(\tau_{\alpha})=0$, $\xi(\tau_{\beta})=\pi/4$,
$\xi(\tau_{\gamma})=3\pi/4$, $\xi(\tau_{\delta})=5\pi/4$
$\xi(\tau_{\epsilon})=7\pi/4$ and $\xi(\tau_{\zeta})=2\pi$,
where $\tau_{\alpha}$ and so on belong to $[\tau_{1},\tau_{2}]$. 
In the intervals $[\tau_{\alpha},\tau_{\beta}]$,
$[\tau_{\gamma},\tau_{\delta}]$ and
$[\tau_{\epsilon},\tau_{\zeta}]$ $\tilde{x}$ cannot be zero if
$T$ is great enough because of (\ref{eq:approximation}) and the
fact that $r(\tau_{0})\geq [(\Nt+\Nth)(\tau_{0})]^{-1/2}$.
For the same reason $\tilde{x}'$ cannot be zero in
$[\tau_{\beta},\tau_{\gamma}]$ and 
$[\tau_{\delta},\tau_{\epsilon}]$ since
$\tilde{x}'=-g\tilde{y}+\epsilon_{x}$. Thus there is exactly one zero
in $[\tau_{\beta},\tau_{\gamma}]$ and one in
$[\tau_{\delta},\tau_{\epsilon}]$ and no zeros in
between. The remaining zeros are picked out in the same way.
By the above it follows that for two consecutive zeros, for example
$\tau_{a}$ and $\tau_{b}$, $\pi/2\leq \xi(\tau_{b})-\xi(\tau_{a})\leq
3\pi/2$. Thus
\[
 |\xi(\tau_{b})-\xi(\tau_{a})-\pi|\leq \frac{\pi}{2}
|\sin(\xi(\tau_{b})-\xi(\tau_{a})-\pi)|\leq
\frac{\pi}{2}(|\cos(\xi(\tau_{a}))|+|\cos(\xi(\tau_{b}))|)
\]
\[
\leq\frac{\pi}{2r(\tau_{0})}(|\tilde{x}(\tau_{a})-x(\tau_{a})|+
|\tilde{x}(\tau_{b})-x(\tau_{b})|)\leq
\frac{9\pi^{2}}{[(\Nt+\Nth)(\tau_{0})]^{1/2}}
\]
\[
\leq\frac{18\pi^{2}}{[(\Nt+\Nth)(\tau_{a})]^{1/2}}
\]
using (\ref{eq:approximation}) and (\ref{eq:nest}). $\Box$

\begin{lemma}\label{lemma:iteration}
Consider a Bianchi VIII solution of
(\ref{eq:whsu})-(\ref{eq:constraint}). There is a $T$ such that
if $\tau_{a}$ and $\tau_{b}$ are two consecutive zeros of $\tx$
and $r(\tau_{a})\geq
[(\Nt+\Nth)(\tau_{a})]^{-1/2}$, $\tau_{a}\geq T$, then if
$z_{1}$ and $z_{2}$ are $\Sp$ evaluated in $\tau_{a}$ and $\tau_{b}$
respectively and $w_{1}$ and $w_{2}$ are $-\No(\Nt+\Nth)$ evaluated in
$\tau_{a}$ and $\tau_{b}$ respectively, we have
\begin{equation}\label{eq:deltasp}
z_{2}-z_{1}=[-(1-z_{1})(1+z_{1})^{2}+\frac{3}{2}w_{1}(2-z_{1})]
(\tau_{b}-\tau_{a})+\epsilon_{1}
\end{equation}
and
\begin{equation}\label{eq:deltan}
w_{2}-w_{1}=(2z_{1}^{2}-2z_{1}+2-3w_{1})w_{1}(\tau_{b}-\tau_{a})+\epsilon_{2}
\end{equation}
where 
\begin{equation}\label{eq:epsest}
|\epsilon_{i}|\leq \frac{C}{(\Nt+\Nth)^{3/2}(\tau_{a})}
\end{equation} 
for $i=1,2$ and some numerical constant $C$.
\end{lemma}
\textit{Remark}. We will prove that $\tau_{b}-\tau_{a}$ is of the
order of magnitude $1/(\Nt+\Nth)$. Consequently we can ignore the
error terms in (\ref{eq:deltasp}) and (\ref{eq:deltan}) as long as 
the polynomial expressions appearing in front of $\tau_{b}-\tau_{a}$
are of the order of magnitude $1$.

\textit{Proof}. The idea is to consider the derivatives of
$\Sp$ and $-\No(\Nt+\Nth)$ and to integrate between $\tau_{a}$
and $\tau_{b}$. We will see that $\Sp$ and $-\No(\Nt+\Nth)$
vary slowly so that we can replace them with constants, up
to an error we estimate. However, $\Sm$ and $\Nt-\Nth$ do not 
vary slowly so that we will have to use (\ref{eq:approximation})
and to estimate $\int_{\tau_{a}}^{\tau_{b}}\sin^{2}(\xi(\tau))
 d\tau$.

All numerical constants below will be denoted $C$.
Observe that
\begin{equation}\label{eq:dtest}
|\tau_{b}-\tau_{a}|\leq\frac{C}{(\Nt+\Nth)(\tau_{a})}
\end{equation}
by the arguments presented in the proof of Lemma \ref{lemma:zeros}, 
assuming $T$ to be great enough.
We will assume $T$ is great enough that $3(\Nt+\Nth)(\tau)\geq 
5(\Nt+\Nth)(\tau_{0})/2 \geq (\Nt+\Nth)(\tau)$ for all 
$\tau\in[\tau_{a},\tau_{b}]$, as is possible by Lemma
\ref{lemma:tnest}. We begin by analyzing the variation
of some relevant expressions in $[\tau_{a},\tau_{b}]$.  
By Lemma \ref{lemma:tnest} we conclude 
\begin{equation}\label{eq:dsp}
|\Sp(\tau)-\Sp(\tau_{a})|\leq \frac{C}{(\Nt+\Nth)(\tau_{a})}
\end{equation}
for all $\tau\in[\tau_{a},\tau_{b}]$ and similarly for
$\No(\Nt+\Nth)$. Assume $\Nt+\Nth$ has a max in $\tau_{\mathrm{max}}$
in $[\tau_{a},\tau_{b}]$ and a min in $\tau_{\mathrm{min}}$ in the
same interval. Then 
\begin{equation}\label{eq:gvar}
 |\frac{1}{(\Nt+\Nth)(\tau_{\mathrm{max}})}-
\frac{1}{(\Nt+\Nth)(\tau_{\mathrm{min}})}|
\end{equation}
\[
\leq\frac{1}{(\Nt+\Nth)(\tau_{\mathrm{max}})}
|1-
\frac{(\Nt+\Nth)(\tau_{\mathrm{max}})}{(\Nt+\Nth)(\tau_{\mathrm{min}})}|
\leq\frac{C}{(\Nt+\Nth)^{2}(\tau_{a})}\nonumber
\]
by (\ref{eq:n23var}). Since $\xi$ as defined in (\ref{eq:xidef}) is
a diffeomorphism we can consider functions of $\tau$ as functions of
$\xi$. We will be interested in the following integral
\[
 \int_{\tau_{a}}^{\tau_{b}}\sin^{2}\xi(\tau) d\tau=
\int_{\xi(\tau_{a})}^{\xi(\tau_{b})}\sin^{2}(\xi)\frac{1}{g} d\xi=
\frac{1}{g(\tau_{a})}\int_{\xi(\tau_{a})}^{\xi(\tau_{a})+\pi}
\sin^{2}(\xi) d\xi
\]
\[
+\frac{1}{g(\tau_{a})}\int_{\xi(\tau_{a})+\pi}^{\xi(\tau_{b})}
\sin^{2}(\xi) d\xi
+\int_{\xi(\tau_{a})}^{\xi(\tau_{b})}\sin^{2}(\xi)\left(\frac{1}{g}
-\frac{1}{g(\tau_{a})}\right)  d\xi
\]
\[
= \frac{\pi}{2g(\tau_{a})}+\delta_{1}
\]
where
\[
|\delta_{1}|\leq \frac{C}{[(\Nt+\Nth)(\tau_{a})]^{3/2}}
\]
due to (\ref{eq:xivar}) and (\ref{eq:gvar}). Furthermore
\begin{equation}\label{eq:deltat}
\tau_{a}-\tau_{b}=\int_{\tau_{a}}^{\tau_{b}} d\tau=
\int_{\xi(\tau_{a})}^{\xi(\tau_{b})}\frac{1}{g} d\xi=
\frac{\pi}{g(\tau_{a})}+\delta_{2},
\end{equation}
by similar arguments, where $\delta_{2}$ is of the same order of 
magnitude as $\delta_{1}$. Consequently
\begin{equation}\label{eq:sinint}
\int_{\tau_{a}}^{\tau_{b}}\sin^{2}\xi(\tau) d\tau=
\frac{1}{2}(\tau_{b}-\tau_{a})+\delta_{3}
\end{equation}
where $\delta_{3}$ is of the same order of magnitude as $\delta_{1}$.
We now have all the necessary estimates at our disposal. In
$[\tau_{a}, \tau_{b}]$ we have
\[
 \Sp'=-(2-q)\Sp-3S_{+}=-(2-2\Sp^2)\Sp+2\Sm^2\Sp-\frac{3}{2}(\Nt-\Nth)^2
+3\No^2
\]
\[
-\frac{3}{2}\No(\Nt+\Nth)=-(2-2z_{1}^{2})z_{1}+
2z_{1}r^{2}(\tau_{a})\cos^{2}(\xi)
-2r^{2}(\tau_{a})\sin^{2}(\xi)
+\frac{3}{2}w_{1}+\epsilon_{3}
\]
where 
\[
|\epsilon_{3}|\leq \frac{C}{(\Nt+\Nth)(\tau_{a})}
\]
due to (\ref{eq:dsp}), the analogous estimate for $\No(\Nt+\Nth)$, 
(\ref{eq:approximation}) and the fact that 
$\No$ is of the order of magnitude $(\Nt+\Nth)^{-1}$
(\ref{eq:constraint}). Integrating, using (\ref{eq:sinint}), we
have
\[
z_{2}-z_{1}=[-(2-2z_{1}^{2})z_{1}+
z_{1}r^{2}(\tau_{a})-r^{2}(\tau_{a})
+\frac{3}{2}w_{1}](\tau_{b}-\tau_{a})+\epsilon_{4}
\]
where
\[
|\epsilon_{4}|\leq \frac{C}{(\Nt+\Nth)^{3/2}(\tau_{a})}.
\]
Using the constraint (\ref{eq:constraint}) to express
$r^{2}(\tau_{a})$ in $z_{1}$ and $w_{1}$ (and $\No$), we deduce
(\ref{eq:deltasp}) and (\ref{eq:epsest}), $i=1$. The argument to 
obtain (\ref{eq:deltan}) is similar. $\Box$

\begin{lemma}\label{lemma:spu}
Consider a Bianchi VIII solution of
(\ref{eq:whsu})-(\ref{eq:constraint}).
For all $\epsilon>0$ there is a $T$ such that $\tau\geq T$ implies
$\Sp(\tau)\leq 1/2+\epsilon$.
\end{lemma}
\textit{Proof}. Let $\epsilon>0$. If $\Sp(\tau)\geq 1/2+
\epsilon/2$
for all $\tau\geq T$, then $-\No(\Nt+\Nth)\rightarrow \infty$
by (\ref{eq:whsu})-(\ref{eq:constraint}) and  Lemma
\ref{lemma:n123l}. Since this cannot 
occur by (\ref{eq:constraint}) we conclude the existence of a
$t\geq T$ such that $\Sp(t)\leq 1/2+\epsilon/2$. If there is 
a $s\geq t$ such that $\Sp(s)\geq 1/2+\epsilon$ there is 
an interval $[\tau_{1},\tau_{2}]$ with $\tau_{1}\geq t$, 
$\Sp(\tau_{1})=1/2+\epsilon/2$, $\Sp(\tau_{2})=1/2+\epsilon$
and $1/2+\epsilon/2\leq\Sp(\tau)\leq 1/2+\epsilon$ for all $\tau
\in [\tau_{1},\tau_{2}]$. Using the constraint (\ref{eq:constraint})
to eliminate $\No(\Nt+\Nth)$ in the expression for $\Sp'$, we conclude
that 
\[
\frac{\epsilon}{2}=\int_{\tau_{1}}^{\tau_{2}}\Sp'(\tau) d\tau\leq
\frac{9}{4}\int_{\tau_{1}}^{\tau_{2}}\No^{2}(\tau) d\tau.
\]
Since $\No'\geq -2\epsilon\No$ in $[\tau_{1},\tau_{2}]$ we conclude
that 
\[
\frac{\epsilon}{2}\leq \frac{9\No^{2}(\tau_{1})}{16\epsilon}.
\]
For $T$ great enough this inequality is impossible
by Lemma \ref{lemma:n123l}. The lemma follows. $\Box$

\begin{lemma}\label{lemma:nnz}
Consider a Bianchi VIII solution of
(\ref{eq:whsu})-(\ref{eq:constraint}).
$\No(\Nt+\Nth)$ does not converge to zero.
\end{lemma}
\textit{Proof}. Assume $\No(\Nt+\Nth)\rightarrow 0$. We prove
that this assumption forces $\Sp$ to become negative which,
in turn, forces $-\No(\Nt+\Nth)$ to increase, leading to a 
contradiction. We use Lemma \ref{lemma:iteration} to 
achieve the decrease in $\Sp$. To apply it we need
$r$ to be big. However, the constraint yields
\begin{equation}\label{eq:r2}
r^{2}=1-\Sp^{2}+\frac{3}{2}\No(\Nt+\Nth)-\frac{3}{4}\No^{2}.
\end{equation}
Since the last two terms converge to zero due to our assumption
and Lemma \ref{lemma:n123l}, only $1-\Sp^{2}$ is of relevance.
From above we have control over $\Sp$ by Lemma \ref{lemma:spu}
and if $\Sp\leq -\frac{1}{2}$ we have nothing to prove as it turns out.

Let us be more precise. Let $T$ be
large enough that $\Sp(\tau)\leq \frac{3}{4}$ for all $\tau\geq T$. This
is possible by Lemma \ref{lemma:spu}. Assume also $T$ to be great
enough that if $t\geq T$ and $-\frac{3}{4}\leq \Sp(t)$, then 
\begin{equation}\label{eq:rest}
r^{2}(t)\geq \frac{1}{4}\geq \frac{1}{(\Nt+\Nth)(t)}.
\end{equation}
In order to achieve this we use (\ref{eq:r2}), Lemma 
\ref{lemma:n123l} and $\No(\Nt+\Nth)\rightarrow 0$. 
Finally let $T$ be great enough that Lemma
\ref{lemma:zeros} and Lemma \ref{lemma:iteration} are applicable 
to all $\tau_{0}\geq T$
such that $r(\tau_{0})\geq (\Nt+\Nth)^{-1/2}(\tau_{0})$.

Let $t\geq T$. We prove that we can iterate $\Sp$ to become 
less than or equal to $-\frac{1}{2}$ using Lemma \ref{lemma:iteration}. 
If $\Sp(t)\leq -\frac{1}{2}$ we are done. The first zero after $t$, 
say $t_{1}$, which exists
by (\ref{eq:rest}) and Lemma \ref{lemma:zeros}, can for the same
reason be assumed to satisfy $\Sp(t_{1})\geq -\frac{1}{2}$. Because of
(\ref{eq:deltasp}), (\ref{eq:epsest}), (\ref{eq:deltat}) and the
fact that $\No(\Nt+\Nth)\rightarrow 0$, we conclude
\begin{equation}\label{eq:desp}
\Sp(t_{2})-\Sp(t_{1})\leq -\frac{1}{9}(t_{2}-t_{1})
\end{equation}
if  $T$ is great enough and $t_{2}$ is
the first zero after $t_{1}$. If $\Sp(t_{k})\geq -\frac{1}{2}$, the next
zero $t_{k+1}$ will satisfy a relation similar to (\ref{eq:desp}).
Sooner or later we will thus have $\Sp(t_{k})\leq -\frac{1}{2}$. Since 
the distance between two zeros becomes arbitrarily small  as time
goes on and the derivative of $\Sp$ is uniformly bounded, $\Sp$ 
cannot at late enough times become greater than $-\frac{1}{4}$ after 
having been smaller than $-\frac{1}{2}$ due to (\ref{eq:desp}). Consequently 
there is a $T'$ such that $\tau\geq T'$ implies $\Sp(\tau)\leq -\frac{1}{4}$. 
We conclude that $-\No(\Nt+\Nth)
\rightarrow \infty$ by considering the derivative of
$\No(\Nt+\Nth)$ and using Lemma \ref{lemma:n123l}. $\Box$

\begin{lemma}\label{lemma:nbig}
Consider a Bianchi VIII solution to (\ref{eq:whsu})-(\ref{eq:constraint}).
For all $\epsilon>0$ there is a $T$ such that $\tau\geq T$ implies
$-\No(\Nt+\Nth)\geq 1/2-\epsilon$. 
\end{lemma}
\textit{Proof}. Assuming $-\No(\Nt+\Nth)\leq 1/2-\epsilon/2$, we
prove that the expression grows at late enough times. Under
certain conditions on $\Sp$ this will be seen by considering the
derivative of $-\No(\Nt+\Nth)$, but under other other circumstances
we apply Lemma \ref{lemma:iteration}. In order for the iteration
to work, we will have to have a good upper bound on $\Sp$, c.f.
(\ref{eq:spu}). We will also need Lemma \ref{lemma:nnz} in order
to get a starting point greater than or equal to some fixed positive
number, irrespective of at how late a time we start.

Let $\frac{1}{10}>\epsilon>0$ and $S$ be such that 
\begin{equation}\label{eq:spu}
\Sp(\tau)\leq \frac{1}{2} +\epsilon^{2}
\end{equation}
for all $\tau\geq S$. 
Define $w=-\No(\Nt+\Nth)$. Since $w$ does not converge to 
zero there is an $\eta>0$ such that for all $S'$ there is a $t\geq
S'$ satisfying $w(t)\geq \eta$. Let $t\geq S$ be such.
We prove that there is a $t'\geq t$ with the property that 
\begin{equation}\label{eq:tprdef}
w(t')\geq \frac{1}{2}-\frac{\epsilon}{2}.
\end{equation}
Assume
\begin{equation}\label{eq:w1ass}
w(t)\leq \frac{1}{2}-\frac{\epsilon}{2}.
\end{equation}

There are two possibilities. Either $\Sp\leq -\frac{1}{4}$, in which case
we will be able to see that $w$ increases by considering its
derivative. If $\Sp\geq -\frac{1}{2}$ we will be able to apply Lemma
\ref{lemma:iteration}.

1. \textit{There is a} $T'$ \textit{such that if} $[t_{1},t_{2}]$ 
\textit{is an interval in which}
$\Sp\leq -\frac{1}{4}$ \textit{and}
$t_{1}\geq T'$, \textit{then}
\begin{equation}\label{eq:cwest}
w(t_{2})-w(t_{1})\geq \frac{1}{2}w(t_{1})(t_{2}-t_{1}).
\end{equation}
This can be seen by applying Lemma \ref{lemma:n123l}, (\ref{eq:whsu})
and (\ref{eq:constraint}). 

2. \textit{There is a} $T''$ \textit{such that if} $t_{1}<t_{2}$ 
\textit{are two consecutive zeros of} 
$\Sm$ \textit{with} $T''\leq t_{1}$, $\Sp(t_{1})\geq -\frac{1}{2}$
\textit{and} $1/2-\epsilon/2\geq w(t_{1})\geq\eta/2$, \textit{then}
\begin{equation}\label{eq:dwest}
w(t_{2})-w(t_{1})\geq
\epsilon w(t_{1})(t_{2}-t_{1}).
\end{equation}
 
We get, in $t_{1}$ 
\begin{equation}\label{eq:r2est}
r^{2}=1-\Sp^2-\frac{3}{4}\No^2+\frac{3}{2}\No(\Nt+\Nth)\geq
\frac{3}{4}\epsilon-3\epsilon^2
\end{equation}
if $T''$ is great enough. 
If $T''$ is great enough we can apply Lemma \ref{lemma:iteration}.
Consider (\ref{eq:deltan}) with 
$\tau_{a}$ replaced with $t_{1}$ and $\tau_{b}$ replaced with
$t_{2}$. Since $2z_{1}^{2}-2z_{1}+2\geq 3/2\ \forall z_{1}\in
\mathbb{R}$ and $1/2-\epsilon/2\geq w(t_{1})$ we get
\[
w(t_{2})-w(t_{1})\geq
\frac{3\epsilon}{2}w(t_{1})(t_{2}-t_{1})+\epsilon_{2}.
\]
Since $w(t_{1})\geq\eta/2$ we can use (\ref{eq:epsest}) and
(\ref{eq:deltat}) to prove that (\ref{eq:dwest}) holds 
if $T''$ is big enough. This is where we need Lemma \ref{lemma:nnz}.

Assume $t$ to be greater than $T'$ and $T''$ and 
that $w(\tau)\leq 1/2-\epsilon/2$ for $\tau\in [t,s]$.
We can divide $[t,s]$ into $t=s_{0}< s_{1}< s_{2}< ... <
s_{k}<s_{k+1}=s$ where $\Sp(s_{i})=-\frac{1}{3}$, $i=1,...,k$ and
$\Sp$ is either $\geq -\frac{1}{2}$ or $\leq -\frac{1}{4}$ in 
each $[s_{i},s_{i+1}]$.
If $1\leq i\leq k$ then  we can change $s_{i}$ so that it becomes
a zero of $\Sm$ without changing the earlier conditions of the
construction except that the new $\Sp(s_{i})$ need not necessarily 
equal $-\frac{1}{3}$, c.f. (\ref{eq:r2est}). This is due to the fact that we
can apply Lemma \ref{lemma:zeros} if $t$ is great enough, that
$|\Sp'|$ is uniformly bounded and the fact that the distance to the
next zero goes to zero as $t$ goes to infinity.  

Since $w(t)\leq 1/2-\epsilon/2$ we can apply Lemma \ref{lemma:zeros}
to get a zero $s_{1/2}$  with $t\leq s_{1/2}$ assuming $t$ is great enough 
and $\Sp(t)\geq -\frac{1}{2}$, c.f. (\ref{eq:r2est}). 
Since $s_{1/2}-t$ is of the 
order of magnitude $1/(\Nt+\Nth)$ and $w'$ is uniformly bounded we 
can assume $w(s_{1/2})\geq \eta/2$. If $\Sp\leq -\frac{1}{4}$ in 
$[s_{0},s_{1}]$, we let $s_{1/2}=t$.
Observe that the condition $w(v)\geq \eta/2$ will be satisfied for a
zero $v$ of $\Sm$ with $s\geq v\geq s_{1/2}$ by 1 and 2. If $\Sp\geq
-\frac{1}{2}$ in $[s_{k},s]$ we have to take into 
account the possibility that $s$ need not be a zero of $\Sm$. Let 
in that case $s_{k+1/2}$ be the last zero of $\Sm$ before $s$. If 
$\Sp\leq -\frac{1}{4}$ in $[s_{k},s]$ we let $s_{k+1/2}=s$. We have,
applying (\ref{eq:dwest}) or (\ref{eq:cwest})  in the
intervals $[s_{i},s_{i+1}]$
\[
w(s_{k+1/2})-w(s_{1/2})\geq \frac{1}{2}\epsilon\eta (s_{k+1/2}-s_{1/2}).
\]
Using the fact that $|w'|$ is uniformly bounded and the estimates
of $s_{1/2}-t$ and $s-s_{k+1/2}$ we conclude that
\[
w(s)-w(t)\geq \frac{1}{2}\epsilon\eta (s-t)-\frac{C}{(\Nt+\Nth)(t)}-
\frac{C}{(\Nt+\Nth)(s)}
\]
for some constant $C>0$. We conclude the existence of a $t'\geq t $
as in (\ref{eq:tprdef}). Finally, $w$ cannot become smaller
than $1/2-\epsilon$ once it has become greater than $1/2-\epsilon/2$
given $t$ great enough. We only have to apply the above and observe
that $|w'|$ is uniformly bounded and the distance between two zeros
of $\Sm$ becomes arbitrarily small as $t\rightarrow \infty$. $\Box$

\begin{thm}\label{thm:b8}
Let $(\Sp,\Sm,\No,\Nt,\Nth)$ be a Bianchi VIII solution to
(\ref{eq:whsu})-(\ref{eq:constraint}) with $\No<0$ and 
$\Nt,\Nth>0$. Then
\[
\lim_{\tau\rightarrow \infty}\No=0,\
\lim_{\tau\rightarrow \infty}\Nt=\infty,\
\lim_{\tau\rightarrow \infty}\Nth=\infty.
\]
Furthermore
\[
\lim_{\tau\rightarrow \infty}\Sp=\frac{1}{2},\
\lim_{\tau\rightarrow \infty}\Sm=0
\]
and 
\[
\lim_{\tau\rightarrow \infty}\No(\Nt+\Nth)=-\frac{1}{2},\
\lim_{\tau\rightarrow \infty}(\Nt-\Nth)=0.
\]
\end{thm}
\textit{Proof}. We begin by proving that for every $\epsilon>0$ there
is a $T$ such that $\tau\geq T$ implies $\Sp(\tau)\geq 1/2-\epsilon$.
We will then be able to draw the conclusions of the theorem. Use the
constraint (\ref{eq:constraint}) to obtain
\begin{equation}\label{eq:sppr}
 \Sp'=-\frac{3}{2}(\Nt-\Nth)^{2}(\Sp+1)+\frac{3}{2}\No(\Nt+\Nth)(2\Sp-1)+
\frac{3}{2}\No^{2}(2-\Sp).
\end{equation}
Observe that by Lemma \ref{lemma:nbig} we can assume $-\No(\Nt+\Nth)$
to be big. By Lemma \ref{lemma:n123l} we can disregard the term
involving $\No$. If $\Sp$ is smaller than, say, $1/2-\epsilon/2$, we
want $\Sp$ to increase. Considering (\ref{eq:sppr}) two things can
happen. Either $r$ is small, in which case the first term is negligible
and then the second term which is positive will dominate. Thus $\Sp$
increases. If $r$ is big, we have to use Lemma \ref{lemma:iteration}.
We make these observations more precise in the following two
statements.

1. \textit{There is a} $T'$ \textit{such that if} 
$r^{2}(\tau)\leq \epsilon/36$ \textit{and} $\Sp(\tau)\leq
1/2-\epsilon/2$ \textit{for} $\tau\in[t,s]$ \textit{where} 
$t\geq T'$, \textit{then} 
\begin{equation}\label{eq:cspest}
\Sp(s)-\Sp(t)\geq \frac{\epsilon}{3}(s-t). 
\end{equation}
The proof is as follows. If $T'$ is great enough
then $-[\No(\Nt+\Nth)](\tau)\geq \frac{1}{3}$ for $\tau\geq T'$ so that if
$\Sp(\tau)\leq 1/2-\epsilon/2$ then
\[
\frac{3}{2}\No(\Nt+\Nth)(2\Sp-1)\geq \frac{\epsilon}{2}
\]
in $\tau$. Assume $T'$ to be great enough that we also have 
\[
\frac{3}{2}[\No^{2}(2-\Sp)](\tau)\leq \frac{\epsilon}{12}
\]
for $\tau\geq T'$. If $\tau\in [t,s]$ we thus conclude that 
\[
\Sp'(\tau)\geq \frac{\epsilon}{3}
\]
using (\ref{eq:sppr}). The statement follows.

2. \textit{There is a} $T''$ \textit{such that if}
$r^{2}(\tau)\geq \epsilon/100$ \textit{and} $\Sp(\tau)\leq
1/2-\epsilon/2$ \textit{for} $\tau\in [\tau_{1},\tau_{2}]$
\textit{where} $\tau_{1}\geq T''$ \textit{and} $\tau_{1}$ 
\textit{and} $\tau_{2}$ \textit{are consecutive zeros of} $\Sm$ 
\textit{then}
\begin{equation}\label{eq:dlsp}
\Sp(\tau_{2})-\Sp(\tau_{1})\geq \frac{\epsilon^{2}}{10}(\tau_{2}-\tau_{1})
\end{equation}

The proof is as follows. We have
\[
r^{2}(\tau)\geq \frac{1}{(\Nt+\Nth)(\tau)}
\]
for $\tau\in[\tau_{1},\tau_{2}]$ if $T''$ is large enough. If
$T''$ is large enough, we can apply Lemma \ref{lemma:iteration}.
Consider (\ref{eq:deltasp}). The relevant polynomial expression
is
\[
 -(1-z_{1})(1+z_{1})^{2}+\frac{3}{2}w_{1}(2-z_{1})=
-(1-z_{1})(1+z_{1})^{2}+\frac{3}{4}(2-z_{1})
\]
\[
+\frac{3}{2}(w_{1}-\frac{1}{2})(2-z_{1})
=(z_{1}+2)(z_{1}-\frac{1}{2})^{2}+
\frac{3}{2}(w_{1}-\frac{1}{2})(2-z_{1}).
\]
Since $-1\leq z_{1}\leq 1/2-\epsilon/2$ in our situation, and 
since we can assume $T''$ to be great enough that 
$w_{1}\geq 1/2-\epsilon^{2}/36$ by Lemma \ref{lemma:nbig}, we
conclude
\[
-(1-z_{1})(1+z_{1})^{2}+\frac{3}{2}w_{1}(2-z_{1})\geq
\frac{\epsilon^{2}}{4}-\frac{\epsilon^{2}}{8}=
\frac{\epsilon^{2}}{8}.
\]
Thus (\ref{eq:dlsp}) follows, in which we have absorbed the error 
term in (\ref{eq:deltasp}) using (\ref{eq:epsest}) and (\ref{eq:deltat}).

Let $S$ be greater than $T'$ and $T''$. We start by proving the 
existence of a $t\geq S$ such that $\Sp(t)\geq 1/2-\epsilon/2$.
Assume $\Sp(\tau)\leq 1/2-\epsilon/2$ in  $[S,s]$. We can divide
the interval by $S=s_{0}<s_{1}<...<s_{k}<s_{n}=s$ where 
$r^{2}$ is either $\geq \epsilon/100$ or $\leq \epsilon/36$ in
each $[s_{i},s_{i+1}]$ and $r^{2}(s_{i})=\epsilon/50$ for all
$s_{i}$ $i=1,...,k$. If $S$ is large enough we can assume the
$s_{i}$ $i=1,...,k$ to be zeros of $\Sm$ without changing the
earlier conditions of the construction other than
$r^{2}(s_{i})=\epsilon/50$, $i=1,...,k$. The reason is that
we can apply Lemma \ref{lemma:zeros} to $s_{i}$, $i=1,...,k$, if
$S$ is large enough, to get a zero within a distance of the
order of magnitude $1/(\Nt+\Nth)$. Using the estimate
(\ref{eq:approximation}) and the fact that $r^{2}=\tx^{2}+\ty^{2}$,
$r^{2}$ can be made to vary an arbitrarily small amount from
the original $s_{i}$ to the first zero after it by choosing $S$
large enough. If $r^{2}(\tau)\geq \epsilon/100$ in $[s_{0},s_{1}]$,
we let $s_{1/2}$ be 
the first zero after $s_{0}$. Otherwise, we let
$s_{1/2}=s$. We define $s_{k+1/2}$ analogously. Using 
(\ref{eq:cspest}) and (\ref{eq:dlsp}) we conclude
\[
\Sp(s_{k+1/2})-\Sp(s_{1/2})\geq \frac{\epsilon^{2}}{10}(s_{k+1/2}-s_{1/2})
\]
assuming $\epsilon<10/3$. If we use our estimates of $s_{1/2}-S$ and
$s-s_{k+1/2}$ we get
\[
\Sp(s)-\Sp(S)\geq \frac{\epsilon^{2}}{10}(s-S)-\frac{C}{(\Nt+\Nth)(S)}- 
\frac{C}{(\Nt+\Nth)(s)}
\]
If $S$ and $s-S$ are large enough this is not possible and there must
thus be a $t$ such that $\Sp(t)\geq 1/2-\epsilon/2$.
 
If $S$ big is enough, $\Sp$ cannot become smaller than $1/2-\epsilon$
once it has been larger than $1/2-\epsilon/2$ due to
(\ref{eq:dlsp}), (\ref{eq:cspest}), the fact that $|\Sp'|$ is
uniformly bounded and the fact that the distance between two zeros
of $\Sm$ goes to zero as $\tau\rightarrow \infty$. 

Combining this observation with Lemma \ref{lemma:spu} we conclude that
\[
\lim_{\tau\rightarrow \infty}\Sp(\tau)=\frac{1}{2}.
\]
The constraint (\ref{eq:constraint}) then yields
\[
 -\No(\Nt+\Nth)\leq\frac{2}{3}(1-\Sp^2-\Sm^2-\frac{3}{4}(\Nt-\Nth)^2-
\frac{3}{4}\No^2)\leq \frac{2}{3}(1-\Sp^2).
\]
Since the right hand side converges to $\frac{1}{2}$ we can use Lemma
\ref{lemma:nbig} to conclude that 
\[
\lim_{\tau\rightarrow \infty}\No(\Nt+\Nth)=-\frac{1}{2}.
\]
By the constraint we can then conclude that 
\[
\lim_{\tau\rightarrow \infty}(\Nt-\Nth)=0
\]
and 
\[
\lim_{\tau\rightarrow \infty}\Sm=0.
\]
$\Box$

\section{Conclusions}\label{section:conclusions}
For all class A spacetimes except IX, $q$ converges to a non-zero 
value as
$\tau\rightarrow\infty$. Thus, by (\ref{eq:reduced}), 
the following theorem follows.
\begin{thm}
For all Bianchi class A vacuum spacetimes except those of type IX, 
the asymptotic behaviour of the reduced Hamiltonian in the expanding
direction is given by
\[
\lim_{\tau\rightarrow \infty}H_{\mathrm{reduced}}(\tau)=0.
\]
\end{thm} 
Consider the metric $\tg$.
By (\ref{eq:rescaled2}) we have 
\[
\tg(\tau)=\sum_{i=1}^{3}\lambda_{i}(\tau)\xi^{i}\otimes \xi^{i}
\]
where 
\[
\lambda_{i}(\tau)=\exp( \int_{0}^{\tau}6\Sigma_{i} d\tau')
\]
by (\ref{eq:lidef}). Note that $\Sigma_{i}$ is a vector determined
by $\Sp$ and $\Sm$ and that the sum of its components is zero. Let us
introduce some terminology.

\begin{definition}
If two $\lambda_{i}$ converge to zero and one to infinity, we say that
the evolution exhibits a \textit{cigar degeneracy}. 
If one $\lambda_{i}$ converges to zero and two to infinity, we call it
a \textit{pancake degeneracy}.
\end{definition}

\subsection{Bianchi I} In this case the $\Sigma_{i}$ are constants satisfying
\[
\sum_{i=1}^{3}\Sigma_{i}^{2}=\frac{2}{3}.
\]
Thus $\lambda_{i}$ converges to $0$, $\infty$ or $1$ depending on the
solution. We get both pancakes and cigars. Concerning the $a_{i}$
in (\ref{eq:aitau}), we observe that if the Bianchi I solution 
corresponds to a special point on the Kasner circle, then two $a_{i}$
are constant and one goes to zero. For all other Bianchi I solutions,
one $a_{i}$ goes to infinity and the other two to zero.

\subsection{Bianchi II}
The $\Sigma_{i}$ converge to $s_{i}$ where the $s_{i}$
satisfy
\[
\sum_{i=1}^{3}s_{i}^{2}=\frac{2}{3}.
\]
The points $(s_{1},s_{2},s_{3})=(\frac{2}{3},-\frac{1}{3},-\frac{1}{3})$ and
the ones obtained by permutating its coordinates are however not allowed as 
limit points (they correspond to the flat Kasner vacuum solutions).
Again we get pancakes or cigars depending on the solution. As far as
the $a_{i}$:s are concerned, one goes to infinity and two go to zero.

\subsection{Bianchi VI${}_{0}$}
In this case the $\Sigma_{i}$ converge to 
$(\frac{2}{3},-\frac{1}{3},-\frac{1}{3})$ or one
of the points obtained by permuting its coordinates.  We get a 
cigar degeneracy. 

Considering the $a_{i}$, (\ref{eq:aitau}) shows that one of the 
$a_{i}$ goes to zero, but what happens to the other two is
unclear.

\begin{prop}
Consider a Bianchi VI${}_{0}$ solution to 
(\ref{eq:whsu})-(\ref{eq:constraint}). Then all the $a_{i}$ converge
to zero as $\tau\rightarrow \infty$.
\end{prop}
\textit{Proof}. Considering (\ref{eq:aitau}) and (\ref{eq:sigi}),
we are interested in the integrals
\[
\int_{0}^{\tau}(1+\Sp\pm\sqrt{3}\Sm) d s=
\frac{1}{2}\int_{0}^{\tau}(2+2\Sp\pm 2\sqrt{3}\Sm) d s.
\]
Observe however that 
\[
(1+\Sp)'=-(2-q)(1+\Sp)
\]
whence
\[
1+\Sp(\tau)=\exp[-\int_{0}^{\tau}(2-q) d s][1+\Sp(0)],
\]
so that, using the expressions for $\Nt'$ and $\Nth'$ in
(\ref{eq:whsu}), we get
\[
a_{2}^{-2}=c_{2}\frac{\Nt}{1+\Sp}
\]
and
\[
a_{3}^{-2}=c_{3}\frac{-\Nth}{1+\Sp}
\]
where $c_{2}$ and $c_{3}$ are positive constants. Consider now the
function
\[
Z_{1}=\frac{\frac{4}{3}\Sm^2+(\Nt-\Nth)^2}{-\Nt\Nth}.
\]
This function is monotone decreasing to the future, and bounded
from below by $2$, since
\[
(\Nt-\Nth)^2\geq -2\Nt\Nth.
\]
Thus it converges to a positive real number, and in consequence, the
same is true of $1/Z_{1}$. Thus, using the constraint,
\[
\frac{-\Nt\Nth}{1+\Sp}=\frac{4}{3}(1-\Sp)\frac{-\Nt\Nth}
{\frac{4}{3}(1-\Sp^2)}=\frac{4}{3}(1-\Sp)\frac{1}{Z_{1}}
\rightarrow \alpha
\]
where $\alpha$ is a positive real number. Since $\Nt$ and $\Nth$
converge to zero as $\tau\rightarrow \infty$ by Proposition
\ref{prop:b6}, we conclude that $a_{2}$ and $a_{3}$ converge
to zero. $\Box$

\subsection{Bianchi VII${}_{0}$}
There are two possibilities. Either the $\Sigma_{i}$ are constants
equal to $(-\frac{2}{3},\frac{1}{3},\frac{1}{3})$, to 
$(\frac{2}{3},-\frac{1}{3},-\frac{1}{3})$ or one of the vectors 
obtained by permuting
the coordinates or they converge to 
$(\frac{2}{3},-\frac{1}{3},-\frac{1}{3})$ (or a permuted
vector). The first possibility yields a pancake or cigar degeneracy. 
The second possibility
yields a cigar degeneracy.

Considering the $a_{i}$ we have the following proposition.

\begin{prop}
Consider a Bianchi VII${}_{0}$ solution to 
(\ref{eq:whsu})-(\ref{eq:constraint}). Then all the $a_{i}$
converge to zero as $\tau\rightarrow \infty$.
\end{prop}
\textit{Proof}. The argument is similar to the Bianchi 
VI${}_{0}$ case, but easier. We just need to observe that
$\Nt,\ \Nth$ are bounded away from zero, and $1+\Sp\rightarrow 0$
by Lemma \ref{lemma:bsz} and Proposition \ref{prop:bsz}.
$\Box$

\subsection{Bianchi VIII}
In this case the $\Sigma_{i}$ converge to 
$(-\frac{1}{3},\frac{1}{6},\frac{1}{6})$ up to
permutations. We have a pancake singularity.

Let us also consider what happens to the $a_{i}$ appearing in 
(\ref{eq:metric}).
Considering (\ref{eq:aitau}), two of the $a_{i}$ converge to zero, 
but what happens to the 
third one is more difficult to say, since the integrand tends to
zero. Here we wish to prove that it is bounded away from infinity.
\begin{prop}
Consider a Bianchi VIII solution to (\ref{eq:whsu})-(\ref{eq:constraint}).
The $a_{i}$ in (\ref{eq:metric}) are bounded as $\tau\rightarrow 
\infty$. 
\end{prop}
\textit{Proof}. As already mentioned two of the $a_{i}$ converge to 
zero. Let us consider the third one. Since $\Sigma_{1}=-2\Sp/3$,
we have
\begin{equation}\label{eq:ao}
a_{1}(\tau)=\exp(\int_{0}^{\tau}(2\Sp-1) d s),
\end{equation}
and in consequence our goal is to prove that $\Sp$ does not get that
much bigger than $\frac{1}{2}$. By Theorem \ref{thm:b8}, we conclude
that $q-4\Sp\rightarrow -\frac{3}{2}$, so that there is a $T_{1}$ such
that
\[
\No(\tau)\leq e^{-\tau}
\]
for all $\tau\geq T_{1}$. Using the constraint (\ref{eq:constraint}) to
eliminate $\No(\Nt+\Nth)$, we deduce
\[
\Sp'=-(1-\Sp^2-\Sm^2)(2\Sp-1)-\frac{9}{4}(\Nt-\Nth)^{2}+
\frac{9}{4}\No^2.
\]
Observe that in the first term, $1-\Sp^2-\Sm^2\rightarrow \frac{3}{4}$.
The term involving $\No^2$ is undesireable, but it can be handled in
the following way. Let
\[
f=\Sp-\frac{1}{2}+\No.
\]
Then
\[
 f'=-(1-\Sp^2-\Sm^2)(2\Sp-1)-\frac{9}{4}(\Nt-\Nth)^{2}+
\frac{9}{4}\No^2+(q-4\Sp)\No.
\]
If $\Sp(\tau)>\frac{1}{2}$, and if $\tau$ is greater than some $T_{2}$,
then we can absorb the term involving $\No^2$ in the term arising from
the derivative of $\No$, and conclude that
\begin{equation}\label{eq:fpr}
f'(\tau)\leq -\alpha f(\tau),
\end{equation}
where $1>\alpha>0$ is a suitable constant. If there is a $T$ such that
$\Sp(\tau)\leq \frac{1}{2}$ for all $\tau\geq T$, we are done, so assume
not. Let us divide the problem
into two subcases. Either there is a $T\geq T_{1},T_{2}$ such that 
$\Sp(\tau)>\frac{1}{2}$ for all $\tau\geq T$, or there is a time
sequence $\tau_{k}\rightarrow \infty$ such that $\Sp(\tau_{k})=\frac{1}{2}$. 
It is convenient to assume $\tau_{1}\geq T_{1},T_{2}$ in this case.
In the first case, we have
\[
f(\tau)\leq f(T)\exp[-\alpha(\tau-T)]
\]
by integrating (\ref{eq:fpr}), and in particular
\[
0\leq \Sp(\tau)-\frac{1}{2}\leq f(T)\exp[-\alpha(\tau-T)]
\]
so that the integral appearing in (\ref{eq:ao}) is finite. 
Consider now the second subcase. Assume $\Sp(\tau)>1/2$ with
$\tau>\tau_{1}$. Let $t<\tau$ be the first point before $\tau$
at which $\Sp(t)=\frac{1}{2}$. Then
\[
f(\tau)\leq f(t)\exp[-\alpha(\tau-t)]\leq
\exp[-t-\alpha(\tau-t)]\leq \exp[-\alpha\tau]
\]
since $\alpha<1$ and $f(t)=\No(t)\leq  e^{-t}$. Thus the 
integral appearing in (\ref{eq:ao})
cannot diverge to infinity. $\Box$

\section*{Acknowledgements}

This research was supported in part by the National Science Foundation
under Grant No. PHY94-07194. Part of this work was carried out
while the author was enjoying the hospitality of the Institute for
Theoretical Physics, Santa Barbara. The author also wishes to 
acknowledge the support of Royal Swedish Academy of Sciences. 
The main motivation for writing this article originated in the
work of Fischer and Moncrief, and the author is grateful to
Lars Andersson for suggesting the problem and Alan Rendall for
suggesting improvements.

\end{document}